\documentclass[a4paper,useAMS,usenatbib]{mn2e}
\usepackage[totalwidth=480pt, totalheight=680pt]{geometry}
\usepackage{graphicx}
\usepackage{aas_macros}
\usepackage{amssymb}   

\newcommand{\pks}{PKS~B1545$-$321 }
\newcommand{\pksns}{PKS~B1545$-$321}

\title[Bow shocks of a relativistic jet?]{PKS~B1545$-$321: Bow shocks of a relativistic jet?}
\author[Safouris, Subrahmanyan, Bicknell \& Saripalli]
{V. Safouris$^{1,2}$\thanks{E-mail: vicky@mso.anu.edu.au (VS)}, R. Subrahmanyan$^{2,3}$,
G. Bicknell$^{1}$ and L. Saripalli$^{2,3}$
\\
$^{1}$Research School of Astronomy and Astrophysics, Mount Stromlo Observatory, 
Australian National University,\\
 Cotter Road, Weston ACT 2611, Australia
\\
$^{2}$Australia Telescope National Facility, CSIRO, Locked Bag 194, 
Narrabri, NSW 2390, Australia
\\
$^{3}$Raman Research Institute, C V Raman Avenue, Sadashivanagar, 
Bangalore 560080, India}

\begin{document}


\pagerange{\pageref{firstpage}--\pageref{lastpage}} \pubyear{2002}

\maketitle

\label{firstpage}

\begin{abstract}
Sensitive, high resolution images of the double-double radio galaxy \pks reveal 
detailed structure, which we interpret in the light of previous work on the 
interaction of restarted jets with pre-existing relict cocoons.  We have also 
examined the spectral and polarization properties of the source, the color 
distribution in the optical host and the environment of this galaxy in order to 
understand its physical evolution. We propose that the restarted jets generate 
narrow bow shocks and that the inner lobes are a mixture of cocoon plasma 
reaccelerated at the bow shock and new jet material reaccelerated at the 
termination shock. The dynamics of the restarted jets implies  that their hot 
spots advance at mildly relativistic speeds with external Mach numbers of at 
least 5. The existence of supersonic hot spot Mach numbers and bright inner 
lobes is the result of entrainment causing a reduction in the sound speed of the 
pre-existing cocoon.   The interruption to jet activity in \pks has been brief 
-- lasting less than a few percent of the lifetime $\sim (0.3-2)\times 10^{8} \> 
\rm yr$ of the giant radio source. The host galaxy is located at the boundary of 
a large scale filamentary structure, and shows blue patches in color 
distribution indicative of a recent merger, which may have triggered the Mpc-
scale radio galaxy.
\end{abstract}


\section{Introduction}
\label{s:intro}

\pks is a remarkable example of a rare type of powerful radio galaxy in which 
\textit{two} double radio structures share the same radio axis and core \citep
{saripalli03a}: \pks is a double-double radio source. It was proposed that the 
inner 300~kpc double source, embedded within the Mpc-scale lobes of the giant 
radio galaxy, represents a new  cycle of activity; that is, newly restarted jets 
have been
`caught in the act' of propagating through the remnant cocoon of a previous 
active phase. There are more than 10 such sources in the literature that have 
been identified as possible restarting radio sources on the basis of the 
detection of inner double structures recessed from the ends of diffuse outer 
lobes.  In the case of \pksns, the outer lobes extend almost all the way to the 
centre without any significant emission gap and the inner double is traced, 
embedded within the outer lobes, all the way from the central core to its bright 
ends. Therefore, \pks is a good candidate for a detailed study of the 
interaction of restarted jets with relict cocoons.  Understanding the physics in 
such interactions is a step towards using the phenomenon of double-double radio 
sources as a probe of the physical nature of extended radio lobes and the causes 
and timescales associated with recurrence in jets in active galactic nuclei.
\par

The existence of double-double radio sources is evidence that the jet activity 
from the central engine of an active galactic nucleus (AGN) is not continuous 
over the life-time of a source: There appear to be interruptions to the flow 
that may be related to either a lack of fuel at the central engine, or 
instabilities in the accretion disk, or instabilities in the jet production 
mechanism. It has been recognized that unless the axis of the central engine 
changes dramatically, the newly restarted jets will initially propagate through 
the relict cocoons of past activity, rather than thermal ambient medium, as in 
the usual case of a single jet outburst. We expect the evolution of the jets to 
be unusual in the cocoon environment, since the energy density in the lobes of a 
radio galaxy is dominated by relativistic magnetized plasma and the cocoon 
medium is much lighter compared to the thermal interstellar and intergalactic 
media. 
\par
Models for the development of jets in synchrotron cocoons have been presented in 
the literature. \citet{clarke91a} and \citet{clarke97a} carried out two-  and 
three-dimensional numerical simulations of an under-dense, supersonic (Mach 6) 
restarting jet, and found that the new jet is heavy with respect to the cocoon 
of the first jet. As a consequence, the restarted jet propagates almost 
ballistically through the pre-existing cocoon, terminating in only a weak (and 
hence not very radio bright) shock. The supersonic advance of the new jet in the 
old cocoon also excites a weak bow shock immediately ahead of the new jet, which 
\citet{clarke97a} suggest should be a visible feature in the radio lobe of a 
restarting radio source. Another feature of the \citet{clarke91a} and \citet
{clarke97a} simulations is that there are no emission features that could be 
ascribed to emission from the cocoon of the restarted jet. This is presumably 
the result of the weak terminal shock combined with the rapid adiabatic 
expansion of the post hot spot flow.  

Whilst these simulations were informative and highlighted aspects of restarting 
jets that were not appreciated at the time (in particular the rapid advance of 
the head of the new jet), they are not consistent with the observations of the 
Mpc-scale double-double radio galaxies of which \pks is an example. Bow shock 
features have not been realized in observations of restarting sources to date: 
The observed inner double sources simply appear as edge-brightened lobes in 
radio images (e.g. \citet{saripalli02a}, \citet{schoenmakers00a}, \citet
{saikia06}). Moreover, the outer ends of the inner doubles have a high contrast 
with respect to the background radio lobes indicating that the terminal shocks 
are at least of moderate strength. (We note however, anticipating observational 
results discussed below, that the inner hot spots are not as bright as one 
observes in normal FR2 radio galaxies.)

A different approach to the observations is to interpret the inner doubles as 
edge-brightened lobes, rather than near-ballistic jets and to invoke thermal 
densities inside the pre-existing cocoons at levels that are higher than 
predicted by simulations. \citet{kaiser00a} proposed that the entrainment of 
dense warm clouds of the intergalactic medium (IGM) could significantly 
contaminate the synchrotron emitting cocoons, so that the development of the new 
jets thereafter would be more consistent with the observations: The interaction 
between the new jet and entrained gas would result in leading hot spots and new, 
filled inner lobes. Additionally, the long timescales for the entrainment and 
dispersion of the contaminating material make this type of model particularly 
relevant to giant radio galaxies, in which inner doubles  are frequently 
observed.  However, leading bow shocks are expected in this model as well. 
\par
Interpreting the inner structures as lobes may solve one problem, namely their 
brightness, but the problem of a missing bow shock remains. 
An additional possibility is that the restarted jets may advance \emph
{hypersonically} with respect to the relict synchrotron lobes of the radio 
galaxy. In this case, the leading bow shock may have a sufficiently small 
opening Mach angle that the emission from both the bow shock and restarted jet 
would together form a narrow inner double structure, appearing as inner lobes 
without any distinct leading bow shock. This possibility was in fact considered 
by \citet{clarke97a} as a possible interpretation of the quasar 3C219. However, 
the required Mach number was of order 1,000. In the case of \pks the required Mach 
number is not nearly so extreme so that this idea is worth pursuing. Other 
variations on this theme include the possibility that changes in direction of 
the restarting jet would widen the working  surface and lead to an apparent 
decrease in the bow shock angle with consequently less stringent requirements on 
the Mach number of the leading hot spot. Another possibility is that the bow 
shock is further influenced by the cocoon matter distribution. We consider these 
possible theoretical interpretations in \S~\ref{s:estimates} and especially \S~
\ref{mach}.

In this paper we present new higher-resolution VLA radio continuum observations 
of the inner double source of PKS B1545$-$321 at 22 and 6~cm wavelengths, and 
examine the detailed structure of the inner lobes in the light of the models 
described above. Our modelling of the propagation and evolution of restarted 
jets in a relict cocoon results in estimates for the entrained thermal densities 
in the outer lobes and the external Mach number of the new jet.  With about five 
synthesized beams across the inner double, our new VLA images of \pks are the 
highest quality (sensitivity and resolution) images to date of any restarting 
source and provide an excellent opportunity for such an investigation.  
Hydrodynamic simulations of restarting jets are being performed in order to 
provide an insight into the appearance of inner doubles in these scenarios; the 
results of these simulations will be presented elsewhere (Safouris et al., in 
preparation).
\par

We adopt a flat cosmology with parameters $\Omega_{\rm 0} = 0.3$, $\Omega_
{\Lambda }= 0.7$ and a Hubble constant $H_{0}=71$ km s$^{-1}$ Mpc$^{-1}$. The 
host galaxy has an absolute magnitude $M_B = -20.9$  and is at a redshift of 
$z=0.1082$ \citep{simpson93}. At this redshift, 1\arcsec = 1.95 kpc.

\section[]{Radio continuum Observations}

Radio observations of the giant radio galaxy \pks were previously carried out 
with the Australia Telescope Compact Array (ATCA) in 2001 \citep{saripalli03a}; 
three separate 12~h Fourier-synthesis observations were done in three different 
array configurations at 22 and 12~cm wavelengths.
\par

We have made new observations of the source with the Very Large Array (VLA) in 
2002, at 22 and 6~cm wavelengths, to improve the imaging sensitivity and 
resolution for the inner source structure: the improved radio continuum images 
form the basis of this work.  A journal of these observations is given in 
Table~1. The source, located at the southern declination of $-32^{\circ}$, is 
visible to the VLA for a total of 6~hr in any single observation. We used the 
hybrid array configurations with the northern extended arm to give near circular 
beams for this southern source. To obtain the maximum resolution, we used the 
BnA array configurations at both observing frequencies. The source was also 
observed at 6~cm with the scaled CnB configuration to obtain visibilities with 
spacings that match the 22~cm BnA data: multi-frequency observations using 
scaled arrays are important for studies of spectral index and polarization as a 
function of wavelength.
\par

In the 22 and 6~cm bands, we used bandwidths of 25 and 50~MHz centered at 1384 
and 4910~MHz respectively. These bandwidths were divided into independent 
channels 3.125 and 12.5~MHz wide respectively, in order to avoid image 
degradation owing to bandwidth smearing.  At 22~cm, we estimate that the 
reduction in peak response at the ends of the $9\arcmin$ outer lobes, as a 
result of bandwidth smearing, is less than 5\%. At 6~cm wavelength, the 
reduction in peak response at the ends of the $3\farcm0$ inner double is also 
less than 5\%.
\par 

The multi-channel continuum visibility data were calibrated using standard 
techniques in AIPS. The channel data were averaged to make a pseudo-continuum 
`channel-0' data-set.  Gains and phases in the `channel-0' visibility data were 
calibrated using interspersed observations of the nearby secondary calibrator 
J1522$-$275. The flux scale was set to the Baars scale \citep{baars77} using an 
observation of 3C286. The observation of 3C286 was also used to calibrate the 
instrumental polarization. The `channel-0' amplitude and phase calibration was 
copied to the multi-channel data and bandpass calibration was determined using 
observations of the secondary calibrator.  
\par

The calibrated 6~cm BnA and CnB visibilities obtained using the VLA were 
concatenated. We constructed images from the 22 and 6~cm VLA visibilities 
separately using MIRIAD, adopting the Clarke algorithm \citep{clark80} to 
deconvolve the strong compact sources and the Steer-Dewdney-Ito algorithm \citep
{steer84} for fainter extended source structure.  Several rounds of phase self-
calibration were iteratively performed.  We combined the self-calibrated VLA 
22~cm visibilities with self-calibrated archival 22~cm ATCA visibilities \citep[from][]
{saripalli03a}; to our knowledge this is the first time that VLA and ATCA 
datasets have been combined and imaged.  We performed a linear mosaic of these 
22~cm VLA and ATCA visibilities using the MIRIAD routine INVERT, which accounts 
for the differing primary beams of the two telescopes. After converting the ATCA 
linear and VLA circular polarization measurements to Stokes parameters, the 
routine images the  two telescope data separately and then combines them in a 
linear mosaic process, where pixels in the individual images are weighted to 
correct for the primary beam attenuation. Deconvolution of the mosaiced 22~cm 
image was done using the maximum entropy algorithm implemented in the MIRIAD 
routine MOSMEM. Our technique for combining and imaging the separate ATCA and 
VLA  observations is described in further detail in Appendix~\ref{s:appendix0}
\par

\subsection{22~cm radio continuum}
\label{s:22cm}
 
The combined VLA and ATCA 22~cm radio image is shown in Figure \ref{vlaatca}. 
The image was made with a beam of FWHM $3\farcs6 \times 2\farcs9$ and has a 
dynamic range of about 200. The rms noise in the image is $\sigma = 40~\mu$Jy 
beam$^{-1}$ and the lowest contour is drawn at $5\sigma$.  The total angular 
extent of the radio source is $8\farcm7$ (corresponding to a projected linear 
size of 1.02~Mpc). From the core, which is coincident with the nucleus of the 
host galaxy, the source is somewhat more extended towards SE; the ratio of the 
extents towards SE and NW is 1.07.  The low-surface-brightness lobes are edge 
brightened and the bridge between the ends is connected with no emission gap. 
There are no hot spots at the ends. The brightest feature in the image is the 2
\farcm9 (340~kpc) inner double structure, which appears to be fully embedded 
within the diffuse lobes of the giant radio galaxy.  
\par

The lobes have a surface brightness that increases towards the ends: the center 
of the bridge has a surface brightness about 100~mJy~arcmin$^{-2}$ rising to 
about 300~mJy~arcmin$^{-2}$ towards the two ends. Relatively bright rims and 
weak emission peaks are observed at the ends, which are just factors of two 
brighter than the surrounding lobes.  These weak emission peaks, or warm spots, 
are well resolved by the $3\farcs6 \times 2\farcs9$ beam and are likely the 
remnants of past hot spots: their presence suggests that the sound crossing 
times in the lobes are not high enough to erase the evidence of the powerful 
jets since they stopped feeding the ends of the outer lobes.  The radio contours 
in Figure \ref{vlaatca}, which are spaced in logarithmic intervals, are tightly 
spaced at the ends indicating a sharp boundary; the relict lobes do not appear 
to have had the time to relax to equilibrium with the surrounding IGM. 
\par 

The total flux density of the \textit{outer} double radio source is 1.7 Jy at 
22~cm, which corresponds to a total radio power of $5\times10^{25}$~W~Hz$^{-1}$ 
at this wavelength. Considering the absolute magnitude of the host galaxy, $M_
{\rm B} = -20.9$, this places the outer lobes of the Mpc-scale radio galaxy in 
the powerful FR-II radio galaxy regime \citep{owen93}, and almost an order of 
magnitude above the FR I/II break in radio power.  This is consistent with the 
edge-brightened morphology of the giant radio lobes.
\par

The inner double structure is clearly seen in the 22~cm radio image as a pair of 
relatively bright 
almost linear features that are entirely contained within the outer lobes of the 
giant radio galaxy. In projection, the inner double and outer lobes have a 
common radio axis and core. The inner source has been traced in the 22~cm image 
all the way from the central core to its bright ends, which are well recessed 
from the outer lobe ends. A gradual curvature is observed over the entire length 
of the inner source, which is reflection symmetric about the core. 
An extrapolation of the inner double structure
shows that the restarted jets are directed towards the brightest warm spots at 
the ends of the relict lobes, indicating that the restarted jets are probably 
tracing the paths of the jets just before the central engine turned off.

\subsection{6~cm radio continuum}
\label{6cm}

The radio image of the inner double source at 6~cm wavelength is shown is Figure 
\ref{inner6}. This image was made with the combined BnA and CnB VLA 
visibilities. The image has an rms noise of 20 $\mu$Jy beam$^{-1}$ and the 
dynamic range is about 125. This high-resolution 6~cm image 
does not have the surface brightness sensitivity to detect the relatively 
fainter parts of the inner lobes closer to the core, which are traced in the 
22~cm image with poorer resolution. The emission trail of the NW inner source 
extends approximately $20\arcsec\>$ from its end towards the core and fades out 
at approximately $55\arcsec\>$ from the core. The emission trail of the SE inner 
source extends twice as far before the surface brightness drops to the same 
detection limit.  The 6~cm images have beem made with a circular beam of FWHM 1
\farcs3; there are $\sim5$ synthesized beams across the width of the inner 
source, and we have well resolved all the complex structure within the length of 
the observed inner source as well as at the ends of the double structure. 
\par
The inner sources have relatively bright ends and lower surface brightness 
emission trailing off towards the core. The emission trails have an almost 
constant deconvolved FWHM about 8\arcsec ($\approx$~16 kpc). The bright 
structures at the leading ends are narrower and have deconvolved FWHM of 3
\arcsec--4\arcsec ($\approx$~6--8 kpc).  As observed projected on the sky, the 
bright ends of the inner structures are at somewhat different distances from the 
core: we measure the arm-length (distance from the core to the ends) ratio to be 
approximately 1.22. 
\par

Our new high resolution 6~cm wavelength VLA image reveals a number of 
characteristics, which 
are not apparent in previous ATCA images of the source. First, patchy emission 
features, which are 2--4 times brighter than the underlying emission, are 
observed along the trail in the SE inner source at 6 cm wavelength (see Figure 
\ref{inner6}). These features are also observed at the same locations in a 22~cm 
wavelength VLA image at 2\farcs6 resolution, which is shown in  Figure~\ref{inner20}, 
indicating that they are genuine. 
\par
The second new feature is that these emission enhancements appear to be located 
on both sides of a central channel along the length of the trail: there appears 
to be a channel running along the center of the SE inner lobe where there is a 
decrement in the surface brightness level. A central decrement was also noted in 
previous 13~cm ATCA images made with a lower resolution \citep{saripalli03a}; 
however, the relatively poorer resolution had led to the inference that the 
inner lobe had the form of an emission sheath whereas with the improved resolution of our new 
6~cm image it is apparent that structure is that of a cylinder with an emission 
decrement along a narrow cylinder running along the axis.  The central decrement 
in the surface brightness is traced all of the way to the head, where it ends at 
the bright peak.  Additionally, our new 6~cm image reveals that the decrement in 
surface brightness follows a {\it straight} line within the southern emission 
trail and, surprisingly, the decrement does not follow the overall curvature 
that is observed over the length of the inner source.  
A mean profile across the width of the SE lobe, which clearly shows the central 
channel, is shown and discussed later in \S~\ref{restartedjets}. We do not 
observe a central decrement in the surface brightness distribution over the 
short length of the northern trail in our VLA 6 cm wavelength image. However, a 
dip in the surface brightness level is detected in profile cuts across the faint 
northern trail in the 22 cm VLA image made with 2\farcs6 FWHM beam 
(see Figure~\ref{inner20}). The decrements trace a straight line from the core to the
 bright peaks at the ends of the inner double. 
\par

The third new feature is that our new VLA 6~cm wavelength image resolves the 
bright ends of the inner double radio source. The image reveals extraordinary 
structures at the leading ends, which were not seen in previous ATCA images. The 
SE inner source terminates in a rim of enhanced emission, which is approximately 
10 times brighter than the trail towards the core. The curved rim extends 
approximately $5\arcsec$ from the tip along the edges of the two sides, and is 
the brightest feature in the radio galaxy, with a peak surface brightness of 0.4 
mJy arcsec$^{-2}$. The NW inner lobe terminates in a well resolved, relatively 
high surface brightness plateau, which is $\sim$4 times brighter than the 
trailing emission. The bright peaks at the ends of the inner sources are sightly 
recessed from the leading edges of the detected emission, at least in 
projection.
\par

The total flux density of the inner double source at 6~cm wavelength is 35~mJy, 
implying that the total radio power of the inner double source is about $2.4
\times10^{24}$~W~Hz$^{-1}$ at 1.4~GHz. The absolute magnitude of the host galaxy 
is $M_{\rm B} = -20.9$ and, according to the relation in \citet{owen93}, the 
total radio power of the inner lobes of the restarting radio galaxy is below the 
FR I-II dividing line. 


\subsection{Polarization and rotation measure}

We have made images of the polarized intensity at 6 and 22 cm wavelengths using, 
respectively, the VLA CnB array visibilities at 6~cm and BnA array visibilities 
at 22~cm. The polarization images, at a resolution of 3\farcs5, were used to 
compute distribution of rotation measure (RM) over the inner double.  We find 
that the RM is fairly uniform, with no significant variations, over each of the 
two components. The NW component has a mean RM of  $-6$~rad~m$^{-2}$ with a 1-$
\sigma$ scatter of 4~rad~m$^{-2}$. The SE inner source has a somewhat lower mean 
RM of $-15$~rad~m$^{-2}$ with a 1-$\sigma$ scatter of 5~rad~m$^{-2}$. The mean 
RM over the entire inner double is  $-11$~rad~m$^{-2}$, which is close to the 
mean value of $-14$~rad~m$^{-2}$ found by \citet{saripalli03a} for the outer 
lobes.
\par
In order to examine the polarization structure in greater detail over the inner 
double, polarization images with beam FWHM 1\farcs3 were made using the BnA and 
CnB visibilities at 6~cm wavelength. In Figure~\ref{pol6}, we show the 
distribution of polarized intensity at 6~cm wavelength. Overlaid are vectors 
showing the orientation of the projected $E$-field, with lengths proportional to 
the fractional polarization. The orientations of the vectors have been corrected 
for Faraday rotation, assuming a uniform $RM = -11$~rad~m$^{-2}$. Peaks in the 
polarized intensity are observed at the ends of the northern and southern 
sources. 
\par

At the leading end of the SE component, there are two peaks in the polarized 
intensity located to the NE and SW of the peak in total intensity. In the trail 
behind the total intensity peak there are two rails of enhanced polarized  
emission. The fractional polarization is enhanced along the boundaries of the 
source and decreases towards the centre-line where there is also a decrement in 
the total intensity. The patchy emission features that were observed in total 
intensity images along the length of the SE component are not coincident with 
peaks in the fractional polarization. The mean level of fractional polarization 
over the SE component is $\sim 25 \%$. Around the rim in the total intensity 
image, where there are peaks in the polarized intensity, the mean fractional 
polarization is $\sim 35 \%$. The projected magnetic field vectors, which are 
perpendicular to the $E$-field orientations, are aligned with the total 
intensity contours around the boundaries of the source.  
\par

The polarization characteristics of the NW component are similar to those of the 
SE component. The mean fractional polarization over its length  is $\sim 25 \%$, 
similar to that seen over the SE component.  The fractional polarization too is 
enhanced along the boundaries. There are several patches where the fractional 
polarization is relatively low; these are located in regions where the 
orientations of the $E$-vectors change sharply in direction, and are likely a 
result of beam depolarization.  Along the boundaries of the northern source, the 
magnetic field is aligned with the total intensity contours. In the central 
regions of the head and trailing emission, the magnetic field vectors are in a 
direction perpendicular to the radio axis.  
\par

We do not find any evidence for Faraday depolarization between 22 and 6~cm 
wavelengths over the NW inner source, for which we measure an average 
depolarization ratio (DR; the ratio between the percentage polarization at 22~cm 
to that at 6~cm) of approximately 1. However, there is an indication of 
depolarization over the SE inner source, where the average DR is approximately 
0.85.  \citet{saripalli03a} estimated the DR in the outer lobes to be in the 
range 0.96--1.0 at the ends and decreasing to approximately 0.92 at the centre, 
with no DR asymmetry between the two lobes.

\subsection{Spectral index}

The distribution of spectral index ($S_\nu \propto \nu^{\alpha}$) over the inner 
double, between 6 and 22 cm wavelengths, was computed using VLA visibilities 
obtained in scaled arrays: CnB data at 6~cm and BnA data at 22~cm. Images at 
both wavelengths were made with beams of FWHM 3\farcs5.  Figure \ref{sindex} 
shows the resulting spectral index distribution; pixels in the individual images 
that had intensities less than $4\times$ the rms image noise were blanked. We 
observe no obvious variations in the distribution of spectral index over the two 
components of the inner double. The NW and SE components of the inner double 
have mean spectral indices of $-0.65\pm0.12$ and $-0.68\pm0.15$ respectively. 
These values are in agreement with the values found by \citet{saripalli03a} 
between 22 and 12~cm wavelengths; therefore, there is no evidence for a 
curvature in the spectra of the two components between 22 and 6 cm  wavelengths.
\par

We have examined the data for trends in the mean index along the length of the 
source by averaging the spectral index distribution shown in Figure \ref{sindex} 
across the width of the source. The resulting profile shows no significant 
trend: we find that along the observable lengths of the inner northern and 
southern sources, from their bright ends to their fading emission trails, the 
spectral index remains fairly constant, and at a value in the range $-0.65$ to 
$-0.70$. 
\par

We have used our new and improved 22~cm image of the source, with the 13~cm 
image presented in \citet{saripalli03a}, to examine the spectral index 
distribution in the outer lobes.
We do not include an image of the computed spectral indices here, 
since our results are consistent with those derived in \citet{saripalli03a}. 
Overall, there is a steepening of the spectral 
index from the outer extremities towards the centre. The spectral indices in the 
warm spots at the ends of the NW and SE outer lobes ($\alpha = -0.67$ and $
\alpha = -0.70$ respectively) are very similar to that in the emission peaks in 
the NW and SE inner lobes ($\alpha = -0.67$). The outer lobe regions with higher 
surface brightness, including those regions adjacent to the peaks at the ends of 
the inner double, have relatively flatter spectral indices.  The lower surface 
brightness bridge in the vicinity of the core and the inner double has a 
spectral index that is relatively steeper and patchy with $\alpha$ in the range 
$-1.0$ to $-2.0$.

\section[]{The host galaxy and environment}
\label{host}

\subsection{B and R band images}
We obtained optical images of the host galaxy of \pks with the Wide Field Imager 
(WFI) at the prime focus of the 3.9~m Anglo-Australian Telescope (AAT) in August 
2004 during service time. We carried out observations in the R-, V- and B-bands  
using one of the eight 2k x 4k CCDs. Conditions were non-photometric with a mean 
seeing of 2\arcsec. The data were reduced using standard techniques in the IRAF 
software package. Frames were bias subtracted and flat-fielded using dome flats. 
Image registration was carried out using stars in the field, whose coordinates 
were measured on the SuperCOSMOS (SCOS) B-band image with an rms accuracy of 0.1 
pixels. Reduced images in the R- and B-bands are shown in Figures~\ref{redwfi} 
and ~\ref{bluewfi}.
\par

The host galaxy appears to be ordinary in the R-band AAT image. Isophotal 
contours, which are shown overlaid on the image, appear to be elliptical and are 
symmetrical about the centre. In contrast, the host galaxy appears somewhat 
disturbed in the B-band image: there is an offset between the centres of the 
inner and outer isophotes and the isophotes are extended to the west. 
Additionally, the B-band image shows a distinct compact object to the SE and 
embedded within the faint envelope of the host galaxy; however, no corresponding 
component is apparent in the R-band image. It was previously noted that the B-
band digitized SuperCosmos image of the host galaxy showed signatures of a 
central dust lane perpendicular to the radio axis \citep{saripalli03a}; however, 
our new higher quality AAT images show no evidence for a dust lane.
\par

We fitted elliptical isophotes to the R- and B-band images using the IRAF 
routine ELLIPSE \citep{jedrzejewski87a}. We ignore the fits to the central $2
\arcsec$ diameter region since they would be compromised by any ellipticity in 
the point spread function. Immediately outside this zone, the fitted elliptical 
isophotes are similar in the red and blue images and have position angles in the 
range 60--80$^{\circ}$. At a distance of 3.6~kpc (1.8~arcseconds) from the 
nucleus, the fitted isophotes abruptly rotate through $>-120^{\circ}$ and 
increase in ellipticity.  In order to reveal any morphological patchiness, we  
constructed models of the host galaxy in the R and B-bands and subtracted these 
from the two AAT images. The IRAF task BMODEL was used to create the models by 
interpolating the elliptical isophotal fits to the data. We did not include the 
third and fourth order harmonics of the elliptical fits in the model subtraction 
as these were found to introduce spurious features in the residual images.  

The residual R- and B-band images are displayed in Figures~\ref{redwfires} and ~
\ref{bluewfires}. Grey-scales in these subtracted images signify excess emission over that expected 
in the models. The R-band residual image does not reveal any peculiar features within the 
brighter regions and envelope of the host galaxy. On the other hand, the B-band 
residual image, clearly reveals a distinct object to the SE of the centre and at 
a projected distance of 10.5~kpc. The counts in this region exceed seven times 
the rms noise in the region where the model has been subtracted. Separately, 
there is a hint of another distinct feature to the west of the nucleus, at a 
level of four times the rms noise. These components, which are relatively blue 
in color, may be associated with star-forming regions in the host galaxy and an 
indication of a recent merger.

\subsection{The neighborhood of \pks from the 6dF Galaxy Survey}
\label{host6df}
In Figure~\ref{6df} we have plotted the positions of all galaxies in the 
vicinity of the host galaxy in redshift space and sky position as measured by 
the 6-degree-Field Galaxy Survey \citep{jones04a,jones05a}, in a set of redshift 
slices.  The 6dF Galaxy Survey (6dFGS) is a near-infrared selected redshift 
survey with median redshift $\bar{z} = 0.05$, complete to $K=12.75$ over the 
entire southern sky ($|b|>10\degr$). The five panels from (a) to (e) each cover 
$40\degr$-fields (280~Mpc) centered on the host galaxy. Each slice is 20~Mpc 
deep in redshift space, and successive redshift slices overlap by 10~Mpc along 
the line of sight.  Panels (a) and (b) together cover a depth of 30~Mpc in front 
of the host galaxy; panel (c) covers a depth of $\pm$10~Mpc about the host 
galaxy, and panels (d) and (e) cover a depth of 30~Mpc behind the host galaxy.  
In panels (a) to (e) of Figure~\ref{6df} we also show in grey scale the local 
galaxy density (per sq.\ deg.) within the individual redshift intervals, 
smoothed by $5\degr$-diameter windows and sampled at cells spaced $1^\circ$ 
apart. Individual 6dFGS fields are $5\fdg7$ in diameter and overlap to varying 
extents. An image showing the completeness of the survey for the sky region is 
also shown in panel (f) of Figure~\ref{6df}: it may be noted here that the 
survey boundary representing the Galactic latitude limit of the survey lies $7
\degr$ from the host galaxy of \pksns.  The effects of redshift incompleteness 
have been corrected for in panels (a) to (e) by weighting the densities using 
the redshift selection function for 6dFGS as a function of sky position (see, 
{\it e.g.}, \citet{jones06a}).  
\par
The host galaxy is not in a high over-density. Fractional over-density
factors up to 8 (which is at the 4 sigma level) are observed within the
280 x 280 x 41 Mpc cube on this smoothing scale. At the location of the
host there are no 6dF galaxies within R=15 Mpc radius; however, the
over-density factors within a few degrees on the sky and within a few
10's of Mpc take on  values up to 3, indicating that the host galaxy is not in
a void either. In velocity space and at redshifts somewhat lower than that of the host,
there appear to be galaxy over-densities to the north and west of the
host.  At redshifts somewhat beyond the host galaxy, the over-densities
are to the south and west.  The distribution is complex and an
elucidation of the detailed structure would require targeted multi-object spectra of
numerous galaxies close to the host on the sky, perhaps using the AAOmega
on the AAT.  On the basis of the distribution in 6dF galaxies, we believe
that the host is located in a relatively low galaxy density environment
in which the fractional over-density is less than 3. It may be noted here
that this result is on a smoothing scale of R=15 Mpc, and the fractional
over-density factor could be higher on smaller scales.
\par
The bottom line is that there is no evidence in the 6dF distribution that
the host is in a rich cluster or relatively dense parts of the large
scale structure of the universe; the host is also not in a void.  We may
conclude that the host is located in the relatively low density parts of
the filamentary galaxy distribution, which has typically fractional
over-densities in the range 5-200.  Assuming that galaxies trace the gas
in the large scale distribution, as is the case in large-scale cosmological 
hydrodynamical simulations \citep{cen99, dave01}, we may expect the IGM 
gas density in the vicinity of the Mpc-scale radio source to be overdense by a 
factor of 3 with respect to the mean baryon density in the IGM associated
with the filaments.  Again, here we are assuming that 
there is no small scale structure in the gas in the vicinity of \pksns. 
Half of the baryons in the universe are missing, presumably in the Warm-Hot IGM 
associated with the filaments \citep{cen99}, and so the gas
density environment of the radio source is at least $3\times 0.5\times 
\Omega_{\rm baryon}\rho_{\rm critical}$ =  $6\times 10^{-31}$ g cm$^{-3}$.

\section[]{Discussion: the phenomenology associated with the restarting jets in 
\pks}

\subsection{The restarted jets}
\label{restartedjets}

On the basis of the Clarke et al. simulations, high resolution images of inner 
doubles in restarting radio galaxies are expected to reveal collimated jet structures, 
with only weak hot spots at their ends. Large-scale jets in powerful radio galaxies are highly collimated \citep{bridle84b}; typically,  they have widths of the order of a few kpc \citep[{\it e.g.}, 3C 353 ][]{swain96a}. However, the ends of the inner double structure in \pks are relatively bright with deconvolved widths of about 7~kpc. Moreover, the trails of edge-brightened emission between the extremities and the core are broader and have deconvolved widths of about 13~kpc.  It is, therefore, unlikely that the entire inner double structure observed in \pks represents collimated restarted jets.
\par

Within the edge-brightened inner double structure we do not observe any 
relatively narrow, collimated, jet-like emission features that may represent 
emission from the restarted jets themselves. We do, however, observe a central 
channel along the length of the southern inner component along which there is a 
decrement in the surface brightness (see \S~\ref{6cm}). A decrement 
is also seen along the length of the short NS trail, excluding the bright end
 (see Figure~\ref{inner20}). These dips in the surface 
brightness may be the signatures of  Doppler dimmed jets. 
 For example, jet material moving with a jet Lorentz factor of $\Gamma_{\rm jet} = 7$ 
 on the plane of the sky would have a Doppler dimming factor of 0.14. Alternatively, 
 the decrement may be the result of the jet being relatively dim 
because the particles in the jet are not reaccelerated and the surrounding 
plasma is freshly accelerated.

In Figure \ref{profile}, we show the mean transverse profile across the southern 
inner source. This slice profile was produced by averaging the trailing emission 
along the length of the source, excluding the bright rim of emission at the end. 
The profile shows clearly the central decrement in surface brightness. 
We have modelled this profile as representing emission from an optically thin, 
thick-walled tube, in which the material in the tube wall has a constant 
emissivity. We convolved the 1-D slice expected from this model with the 
synthesized beam and then fitted the resulting function to the observed profile 
data.  The best fit (also shown in Fig. \ref{profile}) corresponds to an outer 
diameter of 20~kpc for the cylinder and a diameter of 3.2~kpc for the inner 
hollow section.  Our hypothesis is that the central hollow in 
the inner lobes of \pks represents Doppler-dimmed restarting jets.  The 
observations indicate that the inner lobes are not jets, but composed of central 
jets surrounded by inner lobes, which in turn are embedded within the relict outer lobes.
\par
These jets on the two sides of the core appear to be misaligned (i.e. the 
inferred jet paths on each side of the core are not collinear).  Moreover, the jets appear to trace a straight path to the ends of the inner double whereas the inner lobes show mild curvature with a C-shape or reflection symmetry. The structure is not inversion symmetric, as we might expect if the jets are precessing. Presumably, the restarted jets were initially collinear and have gradually evolved to their present misaligned state.

\subsection{Are there bow shocks in the relict cocoon?}
\label{s:bowshocksouter}

Bow shocks are expected to
lead the advance of restarted jets in old cocoons \citep{clarke91a,clarke97a}. We 
expect to observe these shocks via enhanced synchrotron emission since the 
relict lobes of a radio galaxy contain relativistic particles and magnetic 
fields, and the bow shocks should 
enhance the emissivity by compression of the relativistic plasma and 
reacceleration of the electrons. In the synthetic surface brightness images 
derived from the restarting jet simulations, the leading bow shocks do produce 
observable features that are distinct from the relict outer cocoon emission 
\citep{clarke97a}. The bow shock would be expected to weaken with distance from 
the apex, but 
 might remain an observable feature over its entire length. At the apex, where the bow shock 
is stronger and balances the terminal shock of the new jet, the brightness 
contrast might be an order in magnitude or more. In sharp contrast to the predictions of the numerical simulations, bow shocks 
ahead of restarting jets have not been evident in observations of double-double 
radio sources to date \citep[e.g.][]{saripalli02a,saripalli03a}, and this has been 
a major motivation for the high resolution and sensitivity radio observations 
presented here.
\par

In our 22~cm image (Fig.~\ref{vlaatca})
 we measure a mean surface brightness of  450~$\mu$Jy beam$^{-1}$  and 600~$\mu
$Jy beam$^{-1}$ in the relict cocoon emission in the vicinity of the ends of the 
NW and SE inner double; the image rms noise is $40~\mu$Jy~beam$^{-1}$. Within 
the errors in the image, there is no evidence for  wide bow shock like features in the outer relict cocoon  exterior to  the inner double. While there exist regions of enhanced surface brightness around the inner double lobes, notably along the western side of the NW inner 
component, these regions are patchy and may be extensions of the relatively  
brighter material that is observed ahead of the inner double lobes. If there is 
a wider bow shock, which leads the advance of the new jets in the relict lobes, 
then we estimate that the brightness of any such bow shock, relative to the 
cocoon is less than about 20\%. Relative to the hot spots at the ends of the 
restarted jet, the brightness of  such a bow shock is less than 10\%. Our improved images confirm that \pks does not have bow shocks exterior to the inner double and with a brightness expected from simulations \citep[in, \emph{e.g.,}][]{clarke97a}.
\par
However, we note here that the bow shock may be weak over most of its length 
and, therefore, reacceleration of particles in the relict cocoon plasma may be 
insignificant. In this case, any brightness enhancement as a result of the bow 
shock would be the result of compression of the relict lobe plasma.  If the 
relict cocoon emissivity has a straight synchrotron spectrum with spectral index 
$\alpha = -1$, and a tangled magnetic field, 1-D compression by factor $f$ by a 
bow shock would enhance on the emissivity by factor $f^{8/3}$: the limit of 20\% on the enhancement of emissivity, which we have derived above, requires that the compression be at most 7\%.  Curvature in the cocoon electron energy spectrum, which might be 
expected in the relict plasma as a result of enhanced radiative cooling of the 
more energetic particles, may weaken the constraint on the bow shock compression 
ratio. However, the extremely low compression ratio required by the stringent 
observational limits placed by the data presented herein is unlikely.

\subsection{The hot spots at the ends of the inner lobes} 
\label{s:hot spots}

\subsubsection{Total intensity features}
\label{s:intensity_features}

Bright hot spots at the ends of radio lobes represent the working surface where the jet meets the ambient medium. Usually this medium is the relatively dense inter-stellar medium (ISM) or IGM, and the result is a strong shock, which converts a large fraction of the jet kinetic energy into relativistic particles.
The ambient medium of the restarted jets of \pks  are the relic synchrotron lobes of the giant radio source. These are expected to be lighter than the ISM or IGM, even if there has been entrainment of thermal material into the lobes. For this reason, we might expect the appearance of hotspots at the ends of restarted jets to be unusual. In addition, we might also expect that the hotspots at the ends of restarted jets are a merger of both the terminal jet shocks and the bow shocks immediately ahead of the jets since both shocks should reaccelerate relativistic particles and amplify magnetic field. 
\par

The bright ends of the inner double source of \pks are resolved in the 6~cm VLA 
image (Fig. \ref{inner6}). The SE source terminates in a bright extended peak, where as the NW source terminates in a broad plateau of emission. The peak at the end of the SE source 
resembles hot spots created by powerful jets in the IGM; however, the end of the 
NW inner double does not show such a bright compact feature.  \citet{leahy97} 
define a hot spot as any feature that is not part of a jet, that has a largest 
dimension smaller than 10\% of the main axis of the source, a peak brightness 
that is greater then 10 times the rms noise in the image, and that is separated 
from neighboring peaks by a minimum, which drops to less than two-thirds of the 
brightness of the fainter peak. If we adopt this definition, then the peak at 
the tip of the SE inner source may be classified as a hot spot.  By fitting a 
Gaussian to the peak at the end of the jet, we have estimated the intrinsic size 
of the major and minor axes of the central compact feature to be 3\farcs1 and 2
\farcs1 respectively, corresponding to linear sizes of 6.0 and 4.1~kpc. The 
measured size is consistent with the size distribution observed by \citet
{hardcastle98} for a sample of powerful $z<0.3$ radio galaxies. As far as the NW 
structure is concerned it would be overprescriptive to insist that this region 
strictly satisfy the Leahy et al. hot spot criteria. The existence of a  region at 
the NW end of this structure, which is significantly brighter 
than the rest of the tail, is at least consistent with the interaction of a new 
jet with the pre-existing lobe. Also, in view of our interpretation of the arm-length asymmetry of the inner double being due to relativistic motion, this region would be Doppler dimmed (see~\S~\ref{armlength}).
\par

Despite the similarity to classical hot spots, we note that the compact feature 
at the end of the SE inner lobe displays some unusual characteristics that are 
atypical of hot spots observed at the extremities of powerful radio galaxies and 
are, on the other hand, what we might expect of jets developing in a low density 
environment. 
In particular this feature
shows an extended rim of emission, which wraps around the head of the source. 
This rim is extended in a direction perpendicular to the radio axis and shows 
tails, which extend on either side of the central jet (identified to be the 
central decrement). The tails extend approximately 5\arcsec (10~kpc) back toward 
the core of the radio galaxy.  We have examined high resolution images of 
powerful radio galaxies in the literature, and do not find any hot spot that 
resembles the compact structure at the end of the SE inner lobe; no hot spot 
observed to date shows such a neat rim wrapping around an incoming jet. 
Indeed many hot spots project beyond the emision of the associated lobe 
(see \citet{perley97a} and references therein).
The ends of the lobes of powerful radio galaxies often show complex structures, 
and in many cases multiple hot spots. Although there exist examples of multiple 
hot spots distributed in rim-like structures at the ends of sources \citep[{\it 
e.g.}, the northern lobe of 3C173.1,][]{hardcastle97} it is difficult to find 
individual hot spots with swept-back wings or tails. The western lobe of 3C234 
has a central compact component and ridges on each side that bend back into the 
lobe \citep{hardcastle97}; however, in this case the northern ridge is likely to 
be a distinct and separate hot spot. 
We suggest that the extended rim-like structures represents a blending of 
emission from the jet terminal shocks and the immediately preceding bow shock.
\par

The compact feature at the end of the SE inner double is the brightest feature 
in \pksns. However, it has a peak surface brightness of only 0.4~mJy~arcsec$^
{-2}$ at 6~cm wavelength, which is significantly smaller than the typical 
surface brightness observed for hot spots in powerful radio galaxies. For 
example, in the sample of powerful radio galaxies presented in \citet
{hardcastle97} the surface brightness values of the hot spots are an order of 
magnitude or more higher. 
The terminal features at the ends of both the inner SE and NW lobes are 
intermediate in brightness between the strong hot spots observed in FR2 radio 
sources and the low contrast Mach disk structures formed at the ends of the new 
jets in the hydrodynamic restarting jet simulations \citep{clarke91a, 
clarke97a}. We further suggest that this is indicative of jets terminating in an environment 
intermediate in density between the normal IGM and the extremely low density of 
a lobe which has not entrained significant external gas. This assertion is 
treated quantitatively below. 
\par

\subsubsection{Polarization features}
\label{s:pol_features}

The polarization characteristics of the inner components also provide useful 
information.  
At the SE termination, the magnetic field runs around the head of the source, 
neatly following the tails in the total intensity distribution (see Fig.~\ref
{pol6}). At the NW end, the magnetic field is aligned with the total intensity 
contours. In the central regions of the broad plateau the magnetic field is 
transverse to the source axis. The observed polarization properties at the ends 
of the inner northern and southern sources are consistent with what is observed 
in the hot spots of powerful radio galaxies, where the magnetic field is nearly 
always perpendicular to the source axis. Using images with 3\farcs5 FWHM beam, 
we measure the spectral index at the ends of the NW and SE inner sources to be $
\alpha \approx -0.7$ (see Fig.~\ref{sindex}). This is also
consistent with that observed in the hot spots of powerful radio galaxies. 
\par

Notwithstanding the similarities  with the polarization properties of classical 
hot spots there are some intriguing differences.
For example the hot spots in classic FR2 sources such as Pictor~A \citep
{perley97a} exhibit strong polarization coincident with the brightest part of 
the hot spot with the position angle of the E-vector indicating a magnetic field 
perpendicular to the jet direction. In \pks there are local \emph{minima} in the 
polarized intensity either coincident with or nearby to the brightness maxima.  
The directions of the E-vectors along the boundaries of the SE and NW inner 
components in \pks are consistent with the expected alignment of the magnetic 
field along a bow shock. The minima near the positions of peak intensity are 
consistent with the superposition of comparably bright regions in which the 
magnetic field is (a) perpendicular to the jet (the main hot spot emission) and 
(b) aligned almost parallel to the jet (the bow shock emission). There are a 
number of local maxima and minima in the polarization images which could 
possibly be interpreted as the result of jittering of the jet near the 
respective hot spots. However, in view of the straightness of the jets inferred 
from the transverese intensity profiles, we feel that not a great deal of 
jittering is occurring in this region  of the source. On the other hand the 
local maxima could indicate regions where the inner bow-shocked material 
interacts with inhomogeneities in the cocoon.

\section[]{Estimates of source parameters}
\label{s:estimates}

We have noted that there are  several possible restarting jet models, which may 
account for the inner structures observed in \pksns. In the sections 
below, we derive estimates for a number of physical parameters associated with 
the source, including the density of entrained matter in the relict 
cocoon, and the external Mach number of the ends of the restarted jets. 
These estimates allow us to derive some conclusions about the 
formation mechanism of the inner double in this source.

\subsection{Arm length and brightness asymmetry in the inner double}
\label{armlength}

Consideration of the arm length asymmetry in this radio galaxy provides a 
valuable constraint on the physical ideas  discussed in section~\ref{s:intro}. 
One expects the lobes of a powerful radio galaxy to be significantly lighter 
than the ambient ISM or IGM. Hydrodynamical simulations of restarting jets show 
that in case of an interruption, restarted jets are of comparable density to the 
cocoons of the old jets, and as a result they propagate rapidly through the old 
cocoons \citep{clarke91a}. As compared to the original jets in \pksns, we expect 
that the new jets
would propagate relatively unimpeded in their advance from the core, unless 
there has been significant entrainment of thermal material into the relict 
lobes. If the twin oppositely directed jets are advancing at equal and constant 
velocities, then time retardation introduces an asymmetry in their apparent 
projected distances from the core; an argument for constant advance velocities is given 
in section~\ref{s:con}.  In the usual case where the jets terminate in 
the ISM or IGM, caution is required in interpreting arm-length ratios since 
asymmetric distributions of matter around the galaxy can overwhelm time 
retardation effects. However, in \pks the local ISM/IGM has been cleared and the 
new jets are propagating into synchrotron bubbles created by past activity. This 
is probably the most ideal circumstance under which time retardation effects may 
be apparent. 
This would be the case even if there is entrainment of the local IGM, since the 
entrainment rates should be similar on both sides of the source. If there is any 
unevenness in the entrainment rate (as, for example, a result of 
 inhomogeneity in the local IGM distribution), this should be evened out by 
 the dispersal of entrained gas into the large volume of the cocoon. 
\par

Let $d_{\rm head}$ and $d_{\rm c,head}$ be the projected lengths of the ends of the jet 
and counter jet as measured from the core, $\beta_{\rm hs}$ the speed of the hot 
spots relative to the speed of light and $\theta$ the angle of inclination of 
the jet axis to the line of sight. Then the arm length ratio is
\begin{equation}
D = \frac{d_{\rm head}}{d_{\rm c,head}} = \frac{1+\beta_{\rm hs} \cos \theta} 
{1-\beta_{\rm hs} \cos \theta},
\end{equation}
implying that
\begin{equation}
\label{e:bhs1}
\beta_{\rm hs} = \frac{1}{\cos \theta} \frac{D-1}{D+1}.
\end{equation}
The observed arm length ratio for the inner double in \pks is $D=1.22 \pm 0.02$. 
The corresponding values of $\beta_{\rm hs}$ are plotted in Figure~\ref{betahs} 
for angles of inclination in the range $45^\circ < \theta < 90^\circ$. Since 
\pks is a radio galaxy, and not a quasar, we expect that $\theta>45^\circ$.
An immediate implication of equation~\ref{e:bhs1} is that the the angle of 
inclination cannot exceed $\theta = 84\fdg3$, 
since the hot spot speed $\beta_{\rm hs} < 1$.
  The plot shows that for the allowed range of possible viewing angles ($45^
\circ - 84\fdg3$), $\beta_{\rm hs}$ takes on values in the range 0.14--1.0.  The 
lower limit of about $0.14c$ on the advance speed of the new jet is an order of 
magnitude larger than the typical advance speeds inferred for powerful jets 
ploughing into the ISM/IGM \citep{scheuer95a}, consistent with the model that 
the restarted jets in \pks are indeed evolving into a lighter than usual medium.
\par

The brightness asymmetry ratio in hot spots is another consequence of 
relativistic speeds at the ends of jets; the observed ratio of the brightness in 
the hot spots at the NW and SE ends of the inner double is approximately 2.  The 
brightness asymmetry is related to the arm-length asymmetry by
\begin{equation}
R = D^{3-\alpha},
\end{equation}
where $\alpha$ is the spectral index of the hot spots.
Given our observed values of $\alpha=-0.7$ and $D=1.22$, the brightness 
asymmetry is expected to be $R \simeq 2.1$, which is 
appealingly consistent with our measured value of 2.

\par

The measured arm-length asymmetry ratio $D = 1.22$ implies that $\beta_{\rm hs} 
\cos\theta \approx 0.1$. Giant radio sources are selected from compilations of 
large angular size radio sources. Hence,  are biased towards galaxies lying close to the 
sky plane with relatively larger angles of inclination.  This suggests larger 
values for $\beta_{\rm hs}$ . At extreme angles of inclination  where 
$\beta_{\rm hs}$ takes values close to 1, we expect $\beta_{\rm hs} \approx \beta_{\rm jet}$, 
and a ballistic advance of the new jets through the outer synchrotron lobes. This is the case 
predicted by the restarting jet simulations \citep{clarke91a,clarke97a}. 
However, the range of angles for which $\beta_{\rm hs}$ approaches 1 is 
small. For a large range of viewing angles, it is likely that $\beta_{\rm hs} < 
\beta_{\rm jet}$, in which case the advance is not ballistic. This latter scenario is consistent with 
the morphology of the inner lobes and relatively bright hotspots at their ends.

\subsection{Density of entrained matter in the relict cocoon}
\label{rhococoon}

Assuming a non-ballistic advance for the jet ends, we may estimate the density of the 
outer lobes by relating the hot spot pressure to the cocoon density and the hot spot 
advance speed. In this case, the density 
of the cocoon material, $\rho_{\rm c}$, is given by,
\begin{equation}
\label{phs}
\rho_{\rm c} = \frac{3}{4} \frac{p_{\rm hs}}{(\Gamma_{\rm hs} \beta_{\rm hs} c)^
{2}},
\end{equation}
where $p_{\rm hs}$ is the pressure in the hot spot and $\Gamma_{\rm hs}=1
/\sqrt{1-\beta_{\rm hs}^2}$ is the corresponding
Lorentz factor. (This expression does not 
account for all relativistic effects, which we consider below.) 
\par
We estimate a minimum hot spot pressure of $p_{\rm min} = 1.5\times10^
{-11}$ dyne cm$^{-2}$ at the end of the SE inner double, using minimum energy 
conditions \citep{miley80a}. We expect the true hot spot pressure $p_{\rm hs}\ge p_{\rm min}$.
Allowing for relativistic effects, the pressure scales as  $\delta^
{-2}$, where $\delta = \Gamma_{\rm hs}^{-1} (1-\beta_{\rm hs}\cos\theta)^{-1}$ 
is the Doppler factor. Therefore allowing for relativistic effects in the 
minimum pressure estimate, the density in the relict cocoon ahead of the 
restarted jet is:
\begin{equation}
\label{e:rho}
\rho_{\rm c}= \frac{3}{4} \frac{p_{\rm hs}}{p_{\rm min}} 
\frac {p_{\rm min}}{c^2} \frac{(1-\beta_{\rm hs}\cos\theta)^{2}}{\beta_{\rm hs} 
^{2}}.
\end{equation}

The density implied by equation~(\ref{e:rho}) is plotted in Figure~\ref
{rho_theta} as a function of angle of inclination $\theta$ 
for  $p_{\rm hs}/p_{\rm min}$=1. The relationship between $\beta_{\rm 
hs}$ and $\theta$, as implied by the arm-length asymmetry, has been used. The 
plot shows that the cocoon density is in the narrow range $(0.1 - 5) \times 10^
{-31} \> \rm g \> cm^{-3}$ for the range of plausible viewing angles. Repeating the 
calculation for the end of the NW trail, which has a minimum energy pressure 
$p_{\rm min}=1.0\times10^{-11}$  dyne cm$^{-2}$, gives a similar range of density
values. 
\par 

As shown in \S~\ref{host6df},  we expect that $\rho_{\rm IGM}$ is greater than or of the order
 $6\times10^{-31}$~g~cm$^{-3}$. Hence, the cocoon density 
 as a fraction of the IGM density  $\kappa = \rho_{\rm c}/\rho_{\rm IGM}$  is in the range 
 of 0.02 -- 0.8. Smaller values of $\kappa$ are implied for larger angles of inclination.
Our derived values of $\kappa$ are, once again, consistent with the notion that 
the restarted jets in \pks are evolving into a medium that is lighter than the IGM.  At large angles of inclination, the cocoon density 
is small (less than $2\%$ of the IGM density). It is under these conditions that 
we may expect the density of the restarted jets relative to the density of the 
external cocoon ($\eta=\rho_{\rm jet}/\rho_{\rm c}$) to be of order unity or 
greater, as predicted by the restarted jet 
simulations \citep{clarke91a,clarke97a}. Alternatively, at smaller angles of 
inclination, where the cocoon densities represent larger fractions of the IGM 
densities, we expect $\eta<1$ and the entrainment model of \citet
{kaiser00a} to be more relevant. We return to this point again in \S~\ref
{s:gamma_jet}.

\subsection{Expansion timescales for the outer hot spots}
\label{relict}
The jets in the powerful radio galaxy \pks no longer supply energy to the outer 
lobes of the giant source; they have stopped, restarted, and the new jets 
currently terminate at locations that are well recessed from the ends of the 
outer lobes.  As discussed in section~\ref{s:22cm}, the presence of warm spots at the ends of 
the outer lobes suggests that the sound crossing times in the lobes are insifficient to completely dissipate the evidence of the previous jet termination. 
The time required for hot spots with an initial radius $R_{0}$ and pressure $p_
{0}$ to expand to the observed pressure $p_{\rm hs}$ is given by
\begin{equation}
t = \frac{1}{3}\left[ \left(\frac{p_{\rm hs}}{p_{0}}\right)^{-3/4}-1\right] 
\left( \frac{\rho_{\rm c} R_{0}^2}{p_{0}}  
\right)^{1/2}
\end{equation}
(see Appendix~\ref{s:appendix}).

Assuming standard minimum energy conditions \citep{miley80a}, 
we estimate the pressure in the 
remnant hot spots, which have a diameter of $\approx15$ kpc, to be $p_{\rm hs} = 
9\times10^{-13}$~dyne~cm$^{-2}$. Based on the empirical relationship derived by 
\citet{hardcastle98}, we assume that the hot spots at the ends of the 1.05~Mpc powerful giant radio galaxy had a radius of $R_{0} = 2.5$~kpc while the jets were actively feeding them, implying 
(using equation~\ref{e:app}) that the active hot spot pressure was $p_{0} = 8 \times 10^
{-11}$~dyne~cm$^{-2}$. Considering the external cocoon densities derived in section~\ref{rhococoon}, it follows that the jets ceased feeding the hot spots at most $1.8\times10^{5}$~yr ago.

\subsection{The Mach number of the advance of the inner double}
\label{mach}
\subsubsection{Dynamical estimate}
\label{s:M_dynamic}

We may relate the Mach number of advance of the ends of the inner double to the 
speed of the hot spots and the parameters of the synchrotron plasma in the 
relict cocoon traversed by the restarted jets. We use the following 
notation: $\beta_{\rm s, c} = c_{\rm s, c}/c$, where $c_{\rm s, c}$  is the 
sound speed in the relic cocoon; $\Gamma_{\rm s, c}$ is the corresponding 
Lorentz factor \citep{konigl80}. The Mach number of the new hot spots with respect to the sound speed in the relict cocoon is given by
\begin{equation}
M_{\rm hs} = \Gamma_{\rm hs} \beta_{\rm hs}/\Gamma_{\rm s, c} \beta_{\rm s, c}.
\label{e:mach}
\end{equation}

We use the parameter $\chi$ to denote the ratio $\frac{\rho c^2}{4p}$, which parametrizes the rest mass density in a medium where the pressure is dominated by relativistic particles \citep[see][]{bicknell94a}. In the relict cocoon, the density is dominated by thermal matter, and the pressure, $p_{\rm c}$, is dominated by the pressure of the relativistic gas, so that $\chi_{\rm c} =\rho_{\rm c}c^2/4p_{\rm c}$. Similarly, in the jet, $\chi_{\rm jet}=\rho_{\rm jet}c^2/4p_{\rm jet}$. From equation~(\ref{e:rho}) one has

\begin{equation}
\chi_{\rm c} = \frac {3}{16} \, \frac {p_{\rm hs}}{p_{\rm min}} \,
\frac {p_{\rm min}}{p_{\rm c}} \,
\frac {(1 - \beta_{\rm hs}\cos \theta)^2}{\beta_{\rm hs}^2}
\label{e:chi_c}
\end{equation}
In terms of $\chi_{\rm c}$, the sound speed and corresponding 4-velocity in the 
cocoon are given by 
\begin{eqnarray}
\beta_{\rm s, c}^2 &=& 3^{-1}(1 + \chi_{\rm c})^{-1} \\
\Gamma_{\rm s, c} \beta_{\rm s, c} &=& (2+ 3\chi_{\rm c})^{-1/2}
\label{e:cs_4}
\end{eqnarray}
\par
Combining equations~(\ref{e:bhs1}) for the velocity in terms of the arm-length 
asymmetry, (\ref{e:chi_c}) for the density parameter, and equation~(\ref{e:mach}) 
for the hot spot Mach number, enables us to determine the latter as a function of 
the inclination angle.

We can use the above relationships to gain a semi-quantitative idea of the 
implications of the inferred parameters when $\Gamma_{\rm hs }\approx 1$ and $
\Gamma_{\rm s, c}\approx 1$. In this non-relativistic approximation, the 
velocity dependence cancels out and the Mach number of the new hot spot is 
simply given by:
\begin{equation}
M_{\rm hs} \approx \frac{3}{4}\left(\frac{p_{\rm hs}}{p_{\rm c}}\right)^{1/2}
\end{equation}

Assuming standard minimum energy conditions \citep{miley80a}, the relict cocoon 
in the vicinity of the ends of the inner double has a pressure $p_{\rm c} = 3 
\times 10^{-13}$~dyne~cm$^{-2}$.  We may expect the true pressure in the radio 
lobe to be close to that given by the minimum energy condition, since the run of 
$p_{\rm c}$ versus the ratio of particle to magnetic pressures ($p_{\rm part}/
(B^2/8\pi$)) has a shallow minimum, and there is observational evidence that the 
conditions in many radio lobes and even hot spots are close to equipartition 
\citep{croston05a,hardcastle03a}. Hence, using this value for the cocoon 
pressure and the previously derived value for the hot spot pressure, we derive 
an estimate of $M_{\rm hs}\approx5$ for the hot spot Mach number, independent of 
angle of inclination.
\par

Let us now consider the more general case where the ends of the new jets could advance 
with relativistic velocities. As we have seen in section~\ref{armlength}, this is especially
relevant  if the radio axis of the giant source is closely aligned with the 
plane of the sky. In this case, the external Mach number of the hot spot is unbounded as $
\beta_{\rm hs}$ approaches 1. In Figure~\ref{mach_theta}, we show the external hot spot Mach number as a function of viewing angle, as inferred 
from the above equations for $p_{\rm hs}/p_{\rm min} = 1$ and 10.
  These curves demonstrate that for $p_{\rm hs}/p_{\rm min}=1$, hypersonic 
external Mach numbers are only implied for a very narrow range of viewing 
angles. For most viewing angles, the external hot spot Mach number $M_{\rm hs}
\approx 5$, and is only supersonic rather than hypersonic. On the other hand, 
the external Mach number is higher if the pressure in the hot spot is greater 
than the minimum energy estimate.  For example, the $p_{\rm hs} = 10 \, p_{\rm 
min}$ curve shows that over most angles $M_{\rm hs}\approx 15$. Hence the 
inference of a merely supersonic or hypersonic Mach number depends upon the 
ratio of the hot spot pressure to the minimum energy pressure.

\par

\subsubsection{Mach angle estimate}
\label{s:M_angle}

In the hypersonic restarting jet model, the outer edges of the inner lobes in \pks represent the `missing' bow shocks. Thus, it is 
also of interest to derive an estimate of the external Mach number of the new 
hot spots, which is 
based on the opening angles of the putative bow shock.  
\par

If the axis of the inner jet is unchanging and the inner lobe structures 
represent relict lobe material that has been rejuvenated by bow shocks, then the 
widths of the emission trails are representative of the external Mach numbers of 
the restarted jets. The asymptotic angle of the bow shock with respect to the 
jet axis, $\theta_{\rm bs}$, is related to the external Mach number of the hot 
spot by $\theta_{\rm bs} = \sin^{-1} M_{\rm hs}^{-1}$. 

The trail of the inner Southern double is long enough that we can make a 
reasonable estimate of its inclination angle with respect to the radio axis.
In order to measure  the increase in width with distance along the southern 
inner trail, we first rotated the 6~cm image by $-$33\fdg2 to make the jet axis 
vertical and then constructed profiles at different distances from the ends, 
averaging over axial distances of 2.5 synthesized beams in each case. 
 The FWHM of the profiles were measured by including all intensity values 
exceeding 3 times the rms noise. The measured half-widths along the SE inner 
source are plotted in Fig~\ref{f:openang}. The expression for the bow shock angle 
is an asymptotic one. Therefore we did not construct FWHM profiles at the bright 
ends where the width changes abruptly with  length. We did not carry out the same analysis on the NE trail, since it is much shorter and fainter.
In order to measure the opening angles of the low-surface-brightness SE emission 
trail, we least-squares fitted a straight line to the half-width versus distance 
data:  The slope of the fit to the SE inner lobe data implies a half-opening 
angle of 2\fdg5. Simply interpreted, the measured opening angle would imply that 
the SE hot spot is advancing with a Mach number $M_{\rm hs} \approx 23$ with 
respect to the gas in the old cocoon.  
Thus there is a discrepancy between this Mach angle-based estimate of Mach 
number and the dynamically based estimate $\sim 5$ derived above in \S~\ref
{s:M_dynamic}. 

There are at least three possible explanations for this. One 
possibility is that the jet changes direction slightly over its lifetime so that 
the end of the lobe is wider than that associated with a jet which remained 
constant in direction. This would decrease the apparent shock angle. Another 
possible explanation is that the inner lobes are affected by inhomogeneities in 
the cocoon resulting from the entrainment of IGM. In this case, one expects the 
density to increase towards the sides of the coccon and this would have the effect 
of keeping the inner lobes narrow. A third possibility is that the hot spot 
pressure is more than an order of magnitude greater than the minimum energy pressure. 
Given the results of \citet{hardcastle04a} who showed, in a sample of 65 hot 
spots, that departures form equipartition are unlikely, this possibility is 
remote, with the proviso that hot spot dynamics within a pre-existing cocoon 
could be different.

\subsection{The energy budget}
\label{s:ebudget}

The energy budgets corresponding to the original and restarted jets provide 
additional information on the ages and advance speeds of the outer and inner 
radio lobes. 

\subsubsection{The inner lobes}

The FWHM $\approx 3\farcs1$ of the SE hot spot is larger than the jet 
diameter $\approx 1\farcs6$, inferred from the surface brightness decrement
in the SE lobe. This may be the result of the dynamics of 
the shocked plasma near the jet terminus or some wandering about of the jet 
terminus as envisaged in the Dentist Drill model \citep{scheuer82a}. Whatever 
the reason, we take the area $A_{\rm hs}$ of the hot spot as defining the area 
of the working surface.
The balance between the momentum in the jet and that transfered to the relict 
cocoon over $A_{\rm hs}$ leads to the following expression for the energy flux 
in the jet (derived in Appendix~\ref{s:appendix2}):
\begin{equation}
F_{\rm E} = \left[ \frac {1 + 
\frac{\Gamma_{\rm jet}-1}{\Gamma_{\rm jet}} \chi_{\rm jet}}{1 + \chi_{\rm jet}}
\right]
\times \rho_{\rm c} c^3 \beta_{\rm jet} \Gamma_{\rm hs}^{-2}
\frac {(\beta_{\rm hs}/\beta_{\rm jet})^2}{[1 - (\beta_{\rm hs}/\beta_{\rm 
jet})]^2}
A_{\rm hs}.
\label{e:fes}
\end{equation}

The hot spot area $A_{\rm hs} \approx \pi(3\>{\rm kpc})^2$. In Figure~\ref
{fe_theta} we have plotted $F_{\rm E}$ as a function of inclination angle, using 
the arm-length asymmetry relation (equation~\ref{e:bhs1}) and the external 
(cocoon) densities (derived from equation~\ref{e:rho}). 
We have also assumed that $\beta_{\rm jet} \approx 1$.  This plot shows that for 
the plausible range of viewing angles, $F_{\rm E} \ga 10^{44}$~erg~s$^{-1}$; for 
a restricted range of inclination angles $\approx 3.5^\circ$ in which the jet is 
close to the plane of the sky $F_{\rm E}$ might be 1--2 orders of magnitude 
larger than $10^{44}$~erg~s$^{-1}$. However, this is clearly unlikely.
\par

We can check the above estimate of the energy flux against the energy deposited 
in the inner lobes. 
The total energy deposited into the inner lobes by the restarted jets over time 
$\Delta t$ is,
\begin{equation}
U \approx f_{\rm ad} F_{\rm E}\Delta t,
\end{equation} 
where the factor $f_{\rm ad}$ allows for adiabatic losses; we adopt a nominal 
value of $f_{\rm ad} =0.5$, which is typical of a non-thermal bubble inflated by 
a relativistic jet \citep[e.g.][]{bicknell97a}.
\par
The SE inner lobe, which has a projected length of $\approx70$~kpc in our VLA 
6~cm image, has a minimum energy of $U_{\rm min} = 2.8\times10^{57}$~erg.  This 
leads to another expression for the jet energy flux:
\begin{equation}
F_{\rm E} / 10^{44}~{\rm erg~s}^{-1} = 8\beta_{\rm hs}\sin\theta
\label{e:fes2}
\end{equation}
where we have used the relation $\Delta t = l_{\rm obs}/\beta_{\rm hs}c \sin
\theta$ with $\l_{\rm obs} = 70$~kpc. This equation is also plotted as a 
function of inclination angle in Figure~\ref{fe_theta}. The agreement between 
the two independent estimates of the energy flux is excellent for inclination 
angles $\theta<80^{\circ}$. This is independent confirmation of our general assumption 
that the source is not within $10^\circ$ of the plane of the sky.

\subsubsection{The outer lobes}
\label{s:outer_lobes}

The total minimum energy in the outer lobes of the Mpc-scale radio galaxy is $U 
\simeq 6\times10^{59}$~erg.  If we assume that the energy flux of the original 
jets is the same as that for the restarted jets, then we can also estimate the 
active lifetime and speed of advance of the old jets. Adopting $f_{\rm ad} = 1/2
$ and assuming that the original jets also have an energy flux of about $F_{\rm 
E} \approx$~1--6~$\times 10^{44}$~erg~s$^{-1}$  it follows that the old jets 
were active for 0.3--2$ \times 10^{8}$~yr. Assuming that the lobes have not 
expanded significantly in the time since the first jets were switched off, this 
implies an advance speed in the range $\beta_{\rm hs} \approx$ 0.01--0.06$/\sin 
\theta$. 

The advance speed of the original jets is therefore similar to the advance 
velocities inferred for jets in powerful radio sources \citep{scheuer95a}: the 
similarity in advance speeds for giants and smaller size sources suggest that 
the effective hot spot areas are larger in giant sources, where the hot spots 
are farther away from the AGNs, compensating for the lower density media 
encountered by the jets in these sources. 
\par

Finally, we note that if the energy fluxes and efficiencies of the old and 
restarted jets are similar, then 
our analysis of the energy budget is consistent with an order of magnitude 
difference in hot spot speed in the two phases of radio galaxy evolution.

\subsection{Restarting timescales}
\label{timescales}

Suppose the first jet in \pks turns off at $t=t_1$; then the hot spot at the extremity 
of the outer lobe is fed for a further time $L/c \beta_{\rm jet}$, where $L$ is 
the length of the old lobe. The hot spot then expands for a further time $t_{\rm 
exp}$. Hence the current time is
\begin{equation}
t = t_1 + \frac {L}{c \beta_{\rm jet}} + t_{\rm exp}
\end{equation}

Now suppose that the new jet turns on at $t=t_2$ and that the hot spot 
propagates for a distance $l$ at speed $c \beta_{\rm hs}$. This gives  a second 
expression for the current time:
\begin{equation}
t = t_2 + \frac{l}{c \beta_{\rm hs}}
\end{equation}

Hence, equating these two expressions
\begin{equation}
\Delta t = t_2 - t_1 = \frac {L}{c\beta_{\rm jet}} - \frac {l}{c \beta_{\rm hs}} 
+ 
t_{\rm exp}
\label{e:deltat}
\end{equation}
The condition that $t_{2}>t_{1}$ leads to the following condition on $\beta_{\rm 
hs}$,
\begin{equation}
\beta_{\rm hs} > \frac {l}{c t_{\rm exp} + L/\beta_{\rm jet}}
\label{e:beta_hs_lim}
\end{equation}

In \S~\ref{relict}, we computed $t_{\rm exp} \la 1.8 \times 10^5 \> \rm yr$ for the relict hot spots, based on the external lobe densities derived in \S~\ref{rhococoon}. This value, together with $l \approx 150 \> \rm kpc$, $L \approx 490 \> \rm kpc$ (assuming orientation near the plane of the sky) and $\beta_{\rm jet} = 1$, gives $\beta_{\rm hs} \ga 0.27$. This suggests that \pks is observed at an inclination angle of at 
least 70$^\circ$ to the line of sight. In light of this result, together with the unreasonably high energy fluxes implied for extreme inclination angles (see \S~\ref{s:ebudget}), we assume that 
\pks is at an inclination angle $70^{\circ}<\theta<80^{\circ}$, for the 
remainder of the discussion in this paper.

A final point on timescales relates to the time between cessation and renewal of 
jet activity, $\Delta t$. Inserting numbers in equation~(\ref{e:deltat}) for 
$\Delta t$ and taking $\beta_{\rm jet} = 1$ we have
\begin{equation}
\Delta t = 3.2 \times 10^3 \> \left[ 490 - \frac {150}{\beta_{\rm hs}} \right]
+ t_{\rm exp} \>  \rm yr 
\end{equation}
From our estimate of 
$t_{\rm exp} \la 1.8 \times 10^5 \> \rm yr$, it follows that
$\Delta t \approx (5-10) \times 10^5 \> \rm yr$. 
Considering our estimate for the age of the Mpc-scale source in \S~\ref{s:outer_lobes}, this implies that 
$\Delta t$ is at most a few percent of the source age.

\subsection{The jet Lorentz factor}
\label{s:gamma_jet}

In this section we derive estimates of the jet Lorentz factor using 
the above estimates of hot spot velocity.  First, we assume that 
 $\chi_{\rm jet} \sim 1$ 
as inferred in a number of radio jet sources \citep[\emph{e.g.} M87, ][ and Markarian~501, \citeauthor{bicknell01b} 2001]{bicknell96a}. The value of this 
parameter in the cocoon, $\chi_{\rm c}$ takes values between 20--90. This range 
is computed from the range of cocoon densities derived in section~\ref
{rhococoon} for $70^{\circ}<\theta<80^{\circ}$, along with the pressure at 
minimum energy conditions, $p_{\rm c} = 3\times10^{-13}$~dyn~cm$^{-2}$, in the 
external lobe.

For a relativistic jet propagating in a cocoon whose pressure is dominated by 
relativistic plasma, the advance velocity of the hot spots is given by
\begin{equation}
\beta_{\rm hs} \approx \frac {\eta^{1/2} \Gamma_{\rm jet} \beta_{\rm jet}}
{ 1 + \eta^{1/2} \Gamma_{\rm jet}} 
\label{e:bhs}
\end{equation}
where 
\begin{equation}
\eta = \frac {(1 + \chi_{\rm jet})}{(1+ \chi_{\rm c})} \,
\left( \frac {p_{\rm jet}}{p_{\rm c}} \right)
=\left(  
\frac {\rho_{\rm jet} c^2 + 4 p_{\rm jet}}{\rho_{\rm c} c^2 + 4 p_{\rm c}}
\right)
\label{e:eta}
\end{equation}
This equation is equivalent to equation (5) of \citet{marti97}.
 Equation~(\ref{e:bhs}) is relevant when the 
jet is straight and its momentum flux is not spread over a larger area than the 
jet cross-section. This expression is therefore pertinent to the hypersonic jet 
model.

In order to use equation~(\ref{e:bhs}) to determine the jet Lorentz factor, we 
solve for $\Gamma$ to find:
\begin{equation}
 \Gamma_{\rm jet} = \frac {\eta^{1/2} \beta_{\rm hs}^2 + 
 \sqrt{\eta^2(1-\beta_{\rm hs}^2)+\eta \beta_{\rm hs}^2}}
 {\eta(1-\beta_{\rm hs}^2)}
 \label{e:lf}
\end{equation}
In equation~(\ref{e:eta}) for the jet density ratio, we cannot directly measure 
the pressure in the restarted jets. However, we assume that the restarted jets 
are in pressure equilibrium with the inner lobes, which have a minimum energy 
pressure of  $5\times10^{-12}$~dyn~cm$^{-2}$. Thus, using the above estimates, 
we derive values of $\eta$ in the range 0.3--2; this range is consistent with 
the notion that the jets are evolving into a medium of similar density. 
\par

The inferred range of $\eta$ values, together with equation~(\ref{e:lf}) for the 
jet Lorentz factor, imply that $\Gamma_{\rm jet}=$~2--3. Specifically, this if the restarted jets 
advance with hot spot velocities in the range $0.3<\beta_{\rm hs}<0.6$, which 
are relevant to the likely range of inclination angles.


\subsection{Entrainment}
\label{entrainment}

Here we consider the potential origin of 
the gas inside the relict cocoon and show that most of the gas must have been 
entrained. First we estimate the mass density of the non-thermal plasma, $\rho_{\rm nt}$, 
that is deposited by the original jet into the outer lobes. This depends on the 
mass flux of the original jet, $\dot M$, the time for which the first jets were 
active,  $t_{\rm on}$, and the volume of the pre-existing
 cocoon $V_{\rm c}$:
\begin{equation}
\rho_{\rm nt} = \frac{\dot M t_{\rm on}}{V_{\rm c}}.
\end{equation}
\noindent In this expression $\dot M = \rho_{\rm jet}\Gamma_{\rm jet} v_{\rm 
jet} A_{\rm jet}$. 
 
We assume that the original jets have the same mass flux as the restarted jets. 
Using our inferred inner jet diameter $d_{\rm jet} =3.2$~kpc (see \S\ref
{restartedjets}) and an inner jet pressure that is equal to the minimum energy 
inner lobe pressure, $p_{\rm jet}=5\times10^{-12}$~dyn~cm$^{-2}$, 
we estimate $\dot M = 5\times10^{22}\chi_{\rm jet}(\Gamma^{2}-1)^{1/2}$~g~s$^
{-1}$ for the restarted jets, and therefore for the original jets.
\par

Considering the one-sided volume of the outer radio lobes and an active 
timescale of $t_{\rm on}=1\times10^{8}$~yr (see ~\S\ref{s:ebudget}), we obtain 
the following expression for the non-thermal mass density in the outer radio 
lobes,
\begin{equation}
\rho_{\rm nt} \approx 4\times10^{-34}\chi_{\rm jet}(\Gamma_{\rm jet}^{2}-1)^
{1/2}~\rm{g~cm}^{-3}
\end{equation}

Adopting $\chi_{\rm jet} \sim 1$ and $\Gamma_{\rm jet} \sim 2$ gives $\rho_{\rm 
nt} \sim 7\times10^{-34}$~g~cm$^{-3}$. This represents a small fraction (between 
0.7 and 4\%) of the total cocoon density, which is in the range of 
$(2-10)\times10^{-32}\> \rm ~g\>cm^{-3}$, for inclination angles  $70^{\circ}<
\theta<80^{\circ}$. Therefore, a substantial fraction of the mass density in the 
relict cocoon is IGM material ingested into the cocoon at the hot spot, and in 
instabilities along the contact discontinuity.  
At the same time the density in the cocoon is still less than the density of the IGM, which we estimate to be $\sim6\times10^{-31} \rm \> g \> cm^{-3}$.

This calculation highlights the fact that entrainment of the external IGM into 
the outer radio lobes is indeed significant. Consequently, entrainment must play 
an important role in determining the evolution of the restarted jets in the 
relict synchrotron cocoons, as suggested by \citet{kaiser00a} and as required by 
the theoretical interpretation given in this section.

\section{Conclusions}
\label{s:con}

The radio continuum images presented herein of the restarting radio galaxy \pks
show the interaction between the new jets and relict cocoon plasma with the highest 
sensitivity and resolution to date. Most remarkable is the detection of a 
straight 3.2-kpc diameter jet along the axis of the inner SE
 lobe, which has a diameter of 20~kpc. This jet is observed not as an emission 
feature, but as a decrement in total intensity, presumably owing to a Doppler 
dimming of its emission  and/or a lack of accelerated particles prior to the jet plasma 
being shocked at the terminus. 
\par
 We do not detect any evidence for  bow shocks associated with the restarted jets external to the inner lobes in \pksns. The Clarke et al. simulations predict that such features ought to be visible with a brightness comparable to the hotspots. The lack of such expected features, combined with the neat rims of emission at the ends of the inner lobes, lead us to propose that the  
 boundary of the inner double is itself  the bow shock of the restarting jet. Specifically, we suggest that the inner double lobes represent a mixture of relict cocoon material and new jet plasma that has been reaccelerated at the bow shock and terminal jet shocks respectively. It is along these lines that we also predict that there might exist giant radio sources which show no evidence for spectral aging along their axis, and that these sources may have recently restarted and had their entire cocoons reenergized via passage through a bow shock. Indeed, the classic discrepancy between dynamical and spectral ages in radio sources might be due to restarting.
\par
There are a number of observational features that support our above interpretation  of the inner lobes in \pksns:
(1) As mentioned already, 
 we have detected and resolved relatively bright emission peaks 
 at the ends of the inner double
together with rims of emission along the boundary. This morphology is uncharacteristic of classic hot spots which usually protrude beyond the associated lobe. These  rims appear to be consistent with the emission from bow shocks propagating into the background non-thermal plasma. 
(2) The magnetic 
field is aligned with the rims of the inner lobes consistent with the alignment 
of magnetic field in a bow shock. 
(3) Local minima in the polarization of the 
inner lobes are coincident with or adjacent to the brightest features. Again, 
this is uncharacteristic of normal hot spots.  We propose that these features are 
the result of the superimposed emission from moderate strength Mach disks and  
bow shocks, with mutually perpendicular magnetic fields, which we expect from 
supersonic jets terminating in relativistic cocoon material that is light 
compared to the normal IGM.
\par
In addition, we have performed a number of detailed dynamical calculations, which together
 with the new radio continuum images, support the view that the evolution of the new jets in the Mpc-scale lobes is unusual.  We  summarise these as follows and also note that each 
point is consistent  with the above interpretation of the inner lobes as  representing both bow shocks and new jet material:
\begin{enumerate} 
\item The emission peaks at the ends of the inner lobes have a substantially lower
 surface brightness compared to that observed in the hotspots of powerful radio galaxies. 
This is indicative of the jets terminating in a low density environment.
\item Treating the arm length asymmetry in the inner lobes as the result of 
time retardation, we have deduced mildly relativistic hot spot velocities in the 
range $0.3<\beta_{\rm hs}<0.6$ for the inner lobes, consistent with the 
propagation of the jets in a low density environment, but not as low as expected 
from pristine jet material.
\item By considering the hot spot dynamics we have deduced a density in the 
lobes which is less than the likely IGM density but which also exceeds the 
likely density produced by the original jets. The Mach number of the advancing 
hot spot is at least 5 with greater values possible if the minimum energy hot spot 
pressure is significantly exceeded.
\item The hot spot Mach number inferred by associating the spreading rate of the 
SE inner lobe with the Mach angle is approximately 23.
\item Consideration of the timescales related to the expansion of the relict hot 
spots and the propagation of the new hot spots provide a lower limit on the 
advance speed ($\beta_{\rm hs} \ga 0.27$) which is consistent with the range of 
velocities quoted above
\end{enumerate}

The standout discrepancy in the above is between the
two estimates of the Mach number and we have advanced
three possible explanations for this.  The first is that the jet
wanders slightly in its direction producing a broader working
surface thereby decreasing the apparent Mach angle. The
second is that the expansion of the inner lobes is further impeded
by a density gradient toward the sides of the cocoon. The third is that the hot spot 
pressure significantly exceeds the minimum energy value. 
The first idea may imply that the velocity of advance of the jet is not constant 
as we have supposed. It is possible, for example, that the jet evolves according 
to the self-similar $\hbox{velocity} \propto t^{-2/5}$ law for a constant background density in the models of \citet{bicknell97a} and \citet{kaiser97b}. Against this 
is the apparent lack of self-similar {\em spatial} structure in the morphology
of the inner lobes. The self-similar models imply that the ratio of the rate of lateral expansion to forward expansion is constant and the base of the lobes (nearest to the core) should therefore have expanded more than is apparent in the images.
Moreover, unpublished simulations by us of restarting jets show that the new jet is relatively undeflected by the much lower density background medium of the pre-existing lobes. Nevertheless, it is possible that when adequately perturbed at the base such a jet may operate as a ``dentist drill'' but may not propagate as described by the self-similar models; this possibility deserves attention, but detailed investigation is beyond the scope of this paper.

There is support for the second idea that the jet may be affected by local density gradients, in the VLA and ATCA image in Figure~1. The extension of the trailing part of the
NW inner lobe to the west indicates that the northern cocoon
exerts a dynamical influence on the NW inner lobe.
Also, the gradual curvature of the inner lobes points to a
dynamical influence of the cocoon on the lobes. Further simulations
of the interaction of a restarting jet with a pre-existing lobe
should provide some valuable information on these issues.

Notwithstanding this discrepancy between the estimates of the Mach numbers of 
the hot spots, there are several other features of the observations and our 
analysis which are consistent with previous ideas on double-double radio 
sources. The deduction of mildly relativistic hot spot velocities is 
qualitatively consistent with the early models \citep{clarke91a,clarke97a} 
although those models would predict velocities closer to the speed of light and 
fainter inner lobe emission. Therefore we find the notion by \citet{kaiser00a}, 
that entrainment of the surrounding IGM could decrease the sound speed in the 
cocoon, slow down the speed of advance of the inner doubles and make the inner 
lobe emission brighter, to be an appealing one. 
\par
The interruption to jet activity in \pks has been brief. We have shown in \S
\ref{timescales} that given our estimates of cocoon density, it is likely that 
the interruption to jet activity has been no more than a few percent of the age 
of the whole source. This figure could be less than a percent for a source 
lifetime $\sim 2 \times 10^8 \> \rm yr$. 
How likely is it then that we would view this source at this stage in its 
lifetime? Effectively the time available for viewing a restarting jet is given 
by the travel time of the hot spot from the core to the extremity of the source 
$\approx L /c \beta_{\rm hs}$ in the notation of \S~\ref{timescales}. For the 
Northern lobe this is of order $(3-4) \times 10^6 \> \rm yr$ and this could be $
(1.5 - 10)\%$ of the source age, which we have estimated as being $\sim (0.3-2)
\times 10^{8} \> \rm yr$. It is unsurprising therefore that double-double 
sources are relatively rare. Nevertheless, it would be of interest to examine 
the statistics of an unbiased sample of Mpc-scale radio galaxies to see if these 
sorts of fractional time estimates are appropriate.

Finally, we note that we have shown that the host galaxy is located at the 
boundary of a large scale filamentary structure, and shows blue patches in color 
distribution indicative of a recent merger, presumably a minor merger, and it 
may be that the merger event triggered the Mpc-scale radio galaxy. Dating the 
blue stellar populations that we have discovered in the host galaxy would be 
useful in estimating the time-delay between the merger and its influence on the 
emergence of the radio source. However, we should emphasize that the timescales  
we have estimated for the source switching off and then on again are not 
comparable to merger timescales $\sim 10^{8-9} \> \rm yr$ and that one has to 
investigate other processes that would lead to such comparatively small time 
scales.

\begin{table}
\label{tab:params}
\begin{center}
\caption{Journal of VLA observations}
\vspace{6pt}
\begin{tabular}{c c c c}
\hline 
Array       & Frequency  &  Date & Duration  \\
               &        (MHz)       &          & (hr)\\
\hline
BnA     &     1384  & 2002 May 31 &  6 \\
BnA     &     4910  & 2002 Jun  2 & 6 \\
CnB     &     4910  &  2002 Sep 27 &6 \\
\hline 
\end{tabular}
\end{center}
\end{table}

\newpage

\begin{figure*}
\includegraphics[height=\textwidth,angle =-90]{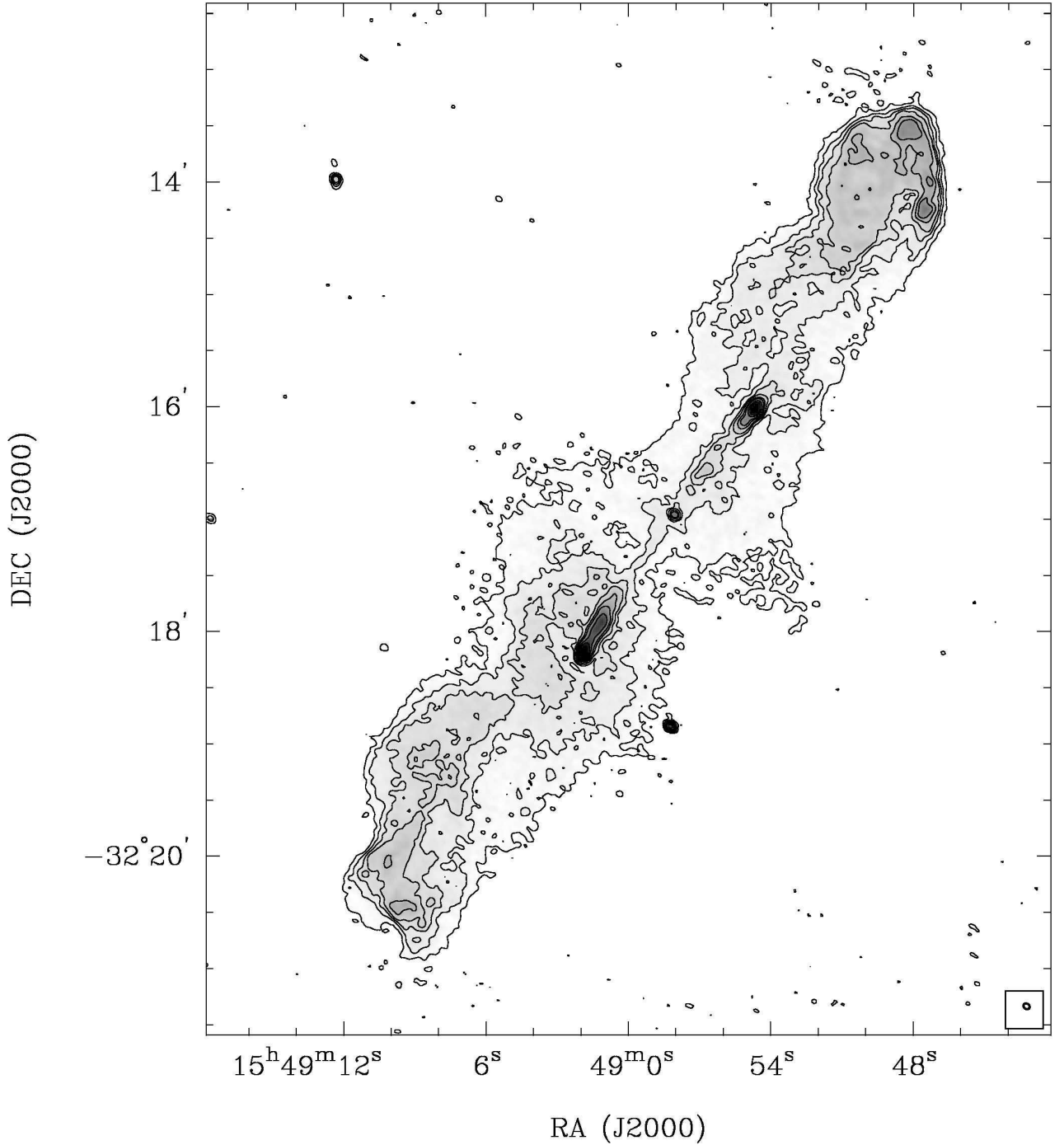}
\caption{\label{vlaatca} The combined VLA and ATCA 22 cm image of PKS
1545$-$321 made with a beam FWHM $3\farcs6 \times 2\farcs9$ at a P.A of 54\degr.
Contours are at (-1, 1, 2, 3, 4, 6, 8, 12, and 16) $\times$ 200 $ \mu$Jy beam$^
{-1}$. 
Grey scales are shown in the range 0.2-4.0 mJy beam$^{-1}$ using a linear scale. 
The rms noise in the image is $40\mu$Jy beam$^{-1}$. The half-power size of the
synthesized beam is displayed in a box in the bottom right-hand corner. This image, as 
well as all others displayed herein, has been corrected for the attenuation due 
to the primary beam.}
\end{figure*}

\newpage

\begin{figure*}
\includegraphics[height=8cm,angle=-90]{f2a.eps}
\includegraphics[height=8cm,angle =-90]{f2b.eps}
\caption{\label{inner6} The southern  and northern inner sources of PKS 1545$-
$321 at 6 cm wavelength made with a beam FWHM $1\farcs3 \times 1\farcs3$. 
Contours are at (-1.0, 1.0, 2.0, 3.0, 4.0, 5.0, 5.5, 6.0, 6.8, 8.0, 9.0, 10.0, 
11.0 and 12.0) $\times$ 65 $\mu$Jy beam$^{-1}$. Grey scales are shown in the 
range 0.065-1.0 mJy beam$^{-1}$  with a linear scale. The rms noise in the image 
is 20 $\mu$Jy beam$^{-1}$.  The half-power size of the synthesized beam is shown 
in the bottom right-hand corner.}
\end{figure*}

\newpage

\begin{figure*}
\includegraphics[height=8cm,angle=-90]{f3a.eps}
\includegraphics[height=8cm,angle =-90]{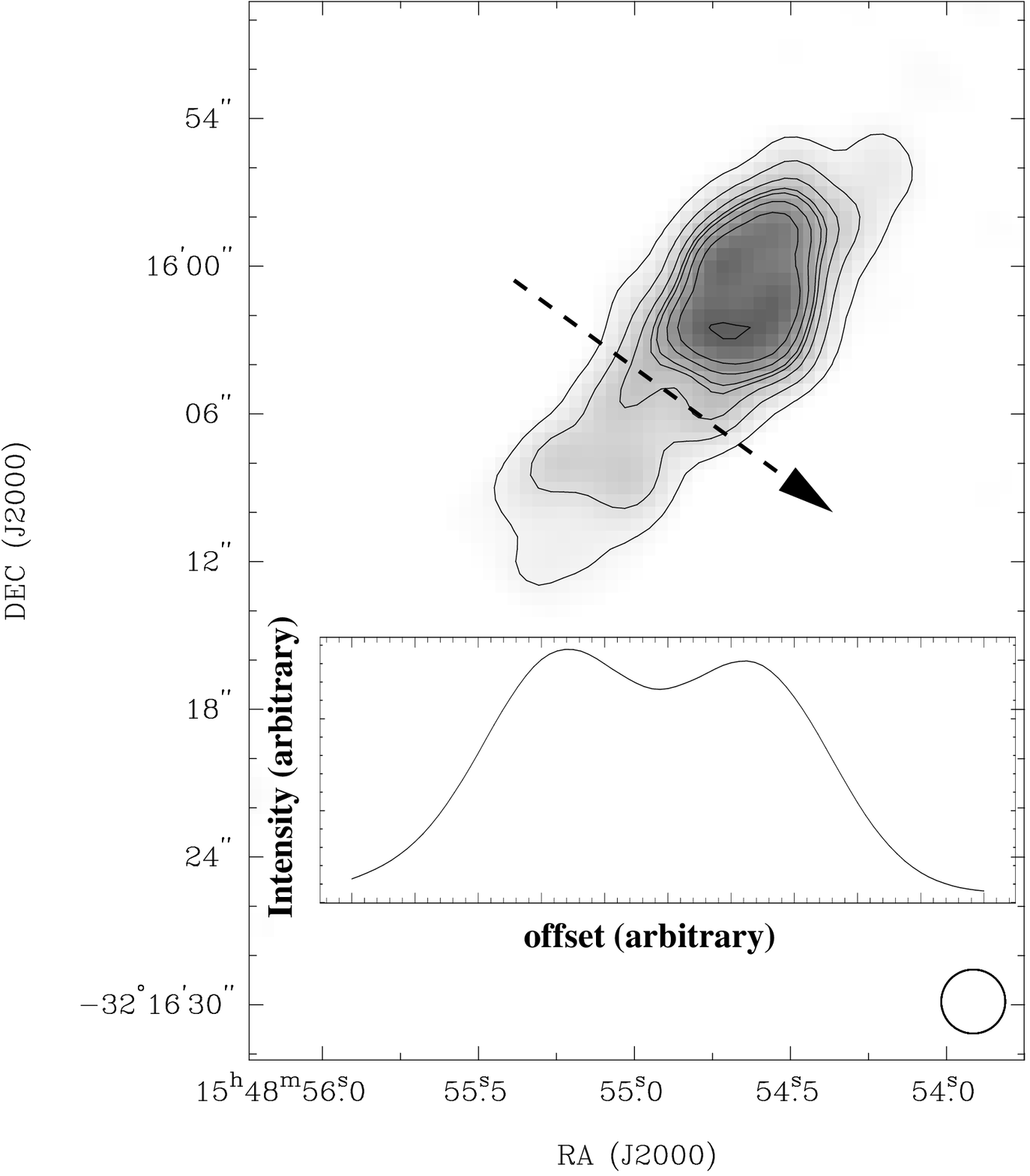}
\caption{\label{inner20} The southern and northern inner sources of PKS 1545$-
$321 at 22 cm wavelength made with a beam FWHM $2\farcs6 \times 2\farcs6$. 
Contours are at (-1.0,1.0,2.0,3.0,3.8,4.2,5.0,6.0,8.0,10.0,12.0) $\times$ 400 $\mu$Jy beam$^{-1}$. Grey scales are shown in the range 0.2-5.0 mJy beam$^{-1}$  with a linear scale. The rms noise in the image 
is 80 $\mu$Jy beam$^{-1}$.  The half-power size of the synthesized beam is shown 
in the bottom right-hand corner. The right-hand panel also shows a slice profile across the NW inner lobe taken along the line indicated by the arrow. }
\end{figure*}

 \newpage

\begin{figure*}
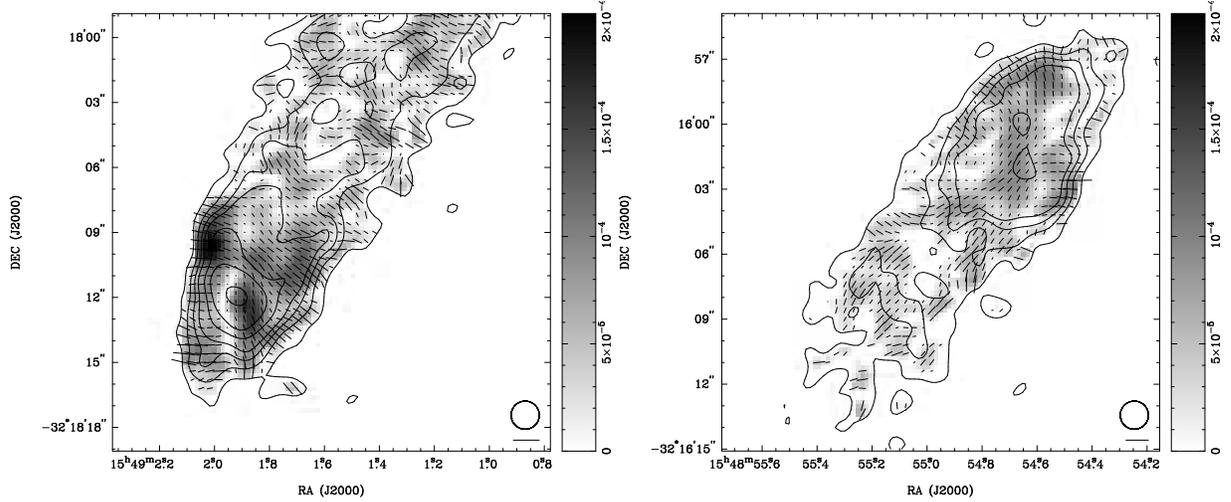

\includegraphics[height=8cm, angle=-90]{f4a.eps}
\includegraphics[height=8cm, angle=-90]{f4b.eps} 
\caption{\label{pol6} Polarization over the southern and northern inner double. 
The images have been made with a beam FWHM 3\farcs5  $\times$ 3\farcs5. The 6 cm 
polarized intensity is shown using grey scales. Bars show the $E$-field 
orientation with lengths proportional to the fractional polarization at 6 cm; 
the vectors shown in the bottom right-hand corners correspond to $100\%$ 
polarization; the orientations of the $E$-vectors have been corrected for the 
line-of-sight Faraday rotation. Contours of the 6 cm total intensity are 
overlaid; contour levels are (-1, 1, 2, 3, 4, 6, 8, and 12) $\times$ 65 $\mu$Jy 
beam$^{-1}$.}
\end{figure*}

\begin{figure*}
\includegraphics[height=15cm, angle=-90]{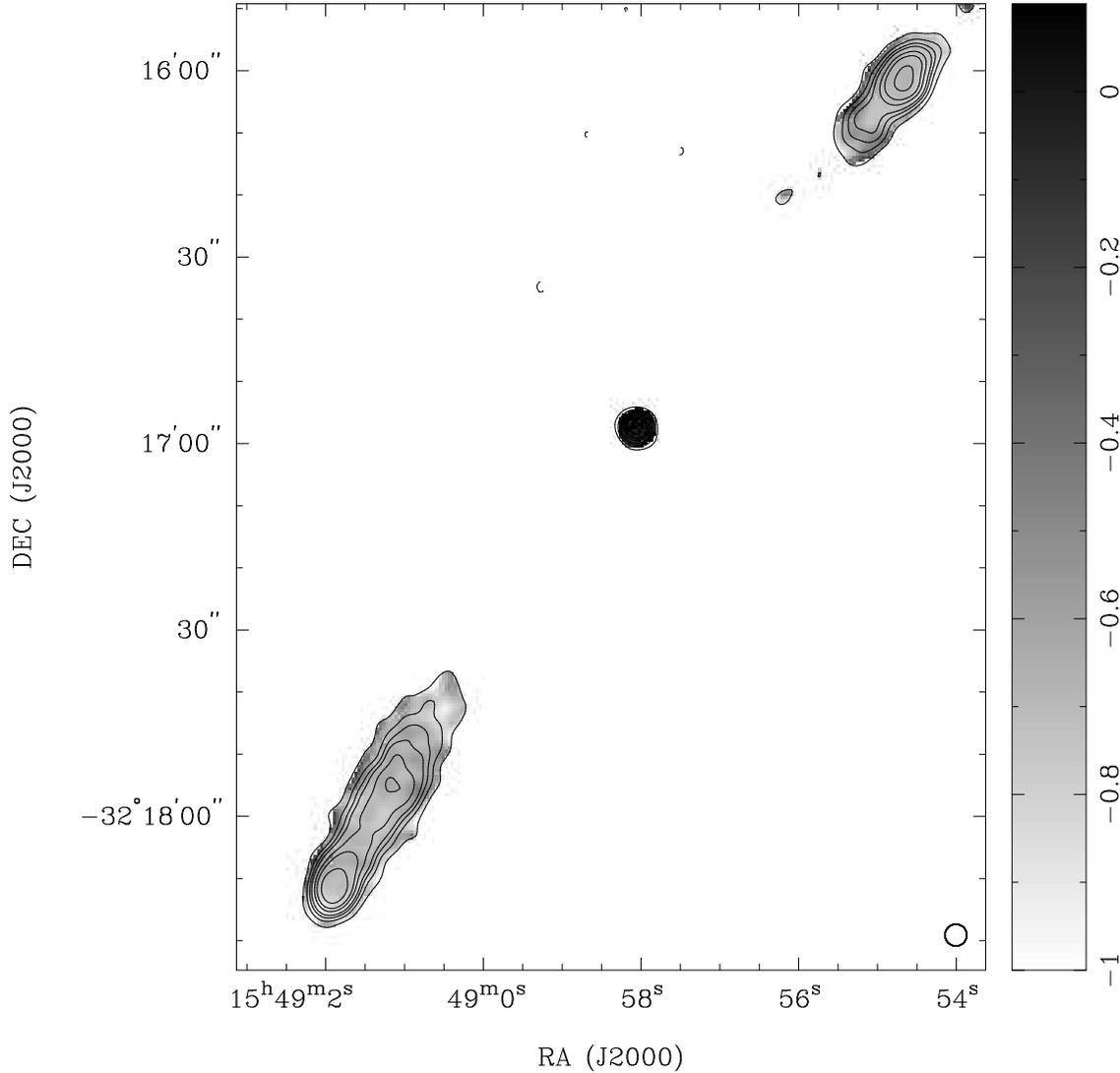}
\caption{\label{sindex} Distribution of spectral index over the inner source 
computed from VLA images at 6 and 22 cm with beams of FWHM $3^{\prime\prime}.5 
\times 3^{\prime\prime}.5$. Grey scales are shown in the range of 0.1 to $-1.0$. 
Contours of the 6 cm total intensity are overlaid; contour levels are (-1, 1, 2, 
3, 4, 6, 8, and 12) $\times$ 160 $\mu$Jy beam$^{-1}$.}
\end{figure*}

\begin{figure*}
\includegraphics[angle=-90, width=10cm]{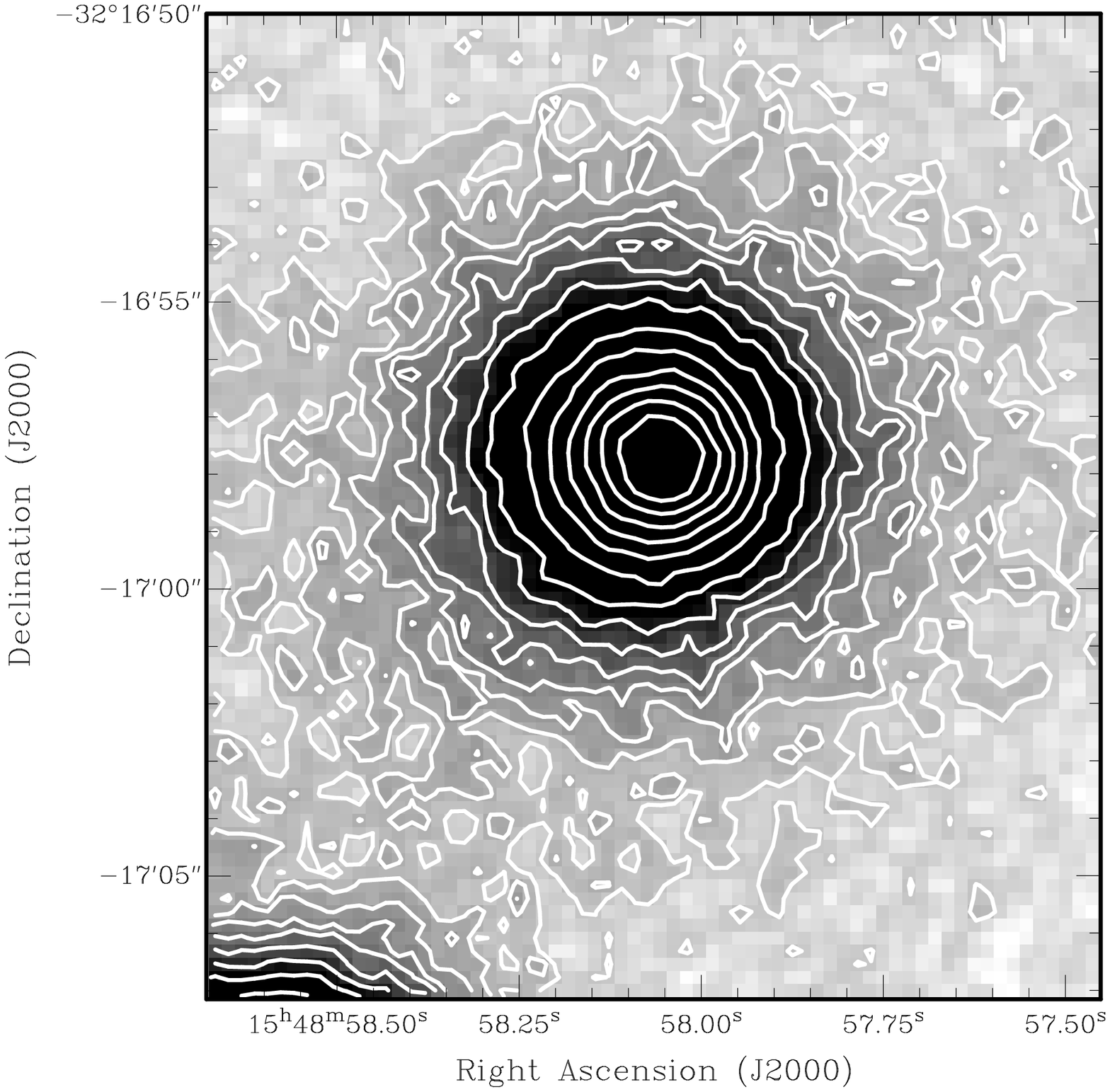}
\caption{\label{redwfi} R-band AAT WFI image of the host galaxy.}
\end{figure*}
\begin{figure*}
\includegraphics[angle=-90,width=10cm]{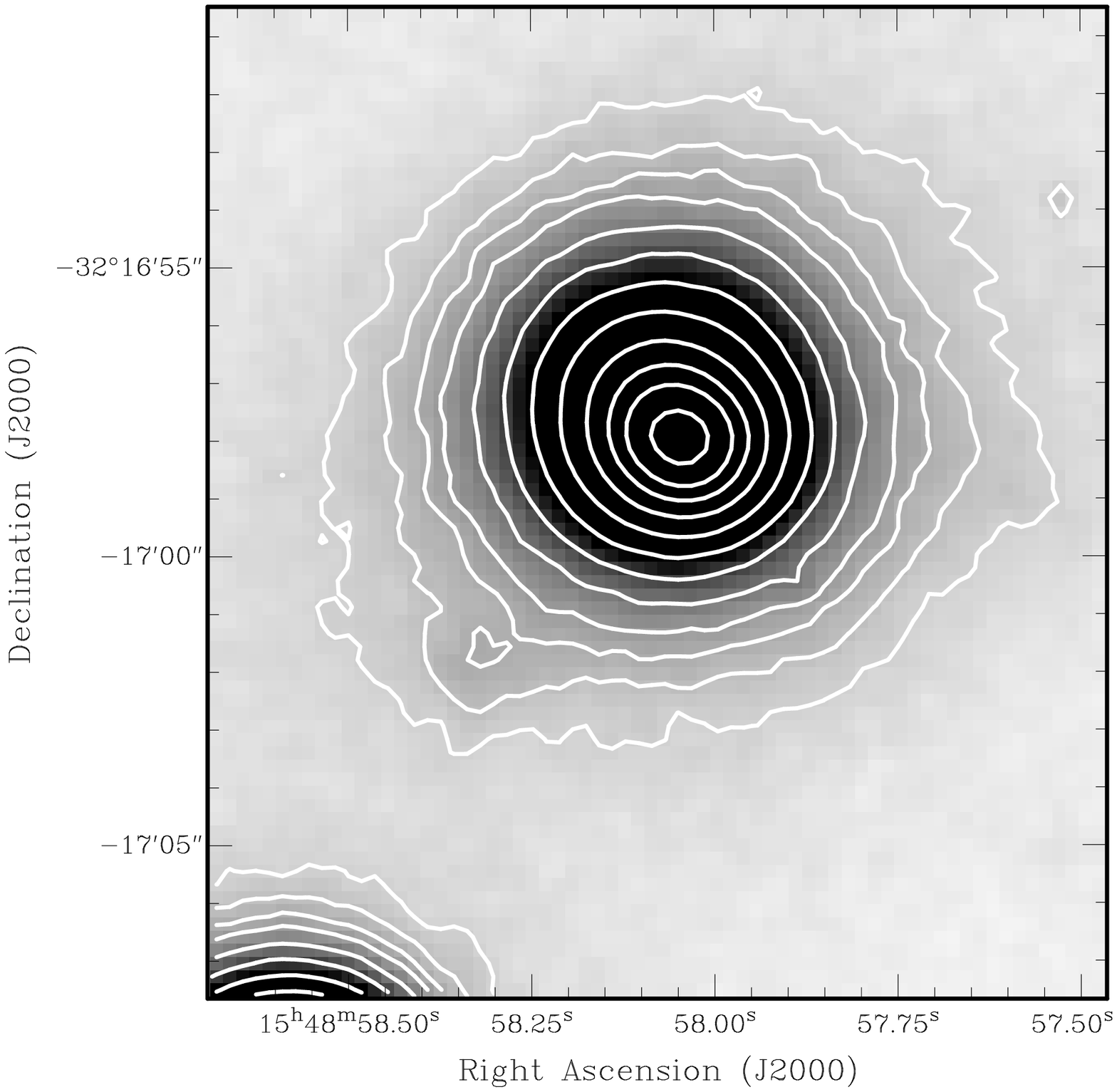}
\caption{\label{bluewfi} B-band AAT WFI image of the host galaxy. }
\end{figure*}

\begin{figure*}
\includegraphics[angle=-90, width=10cm]{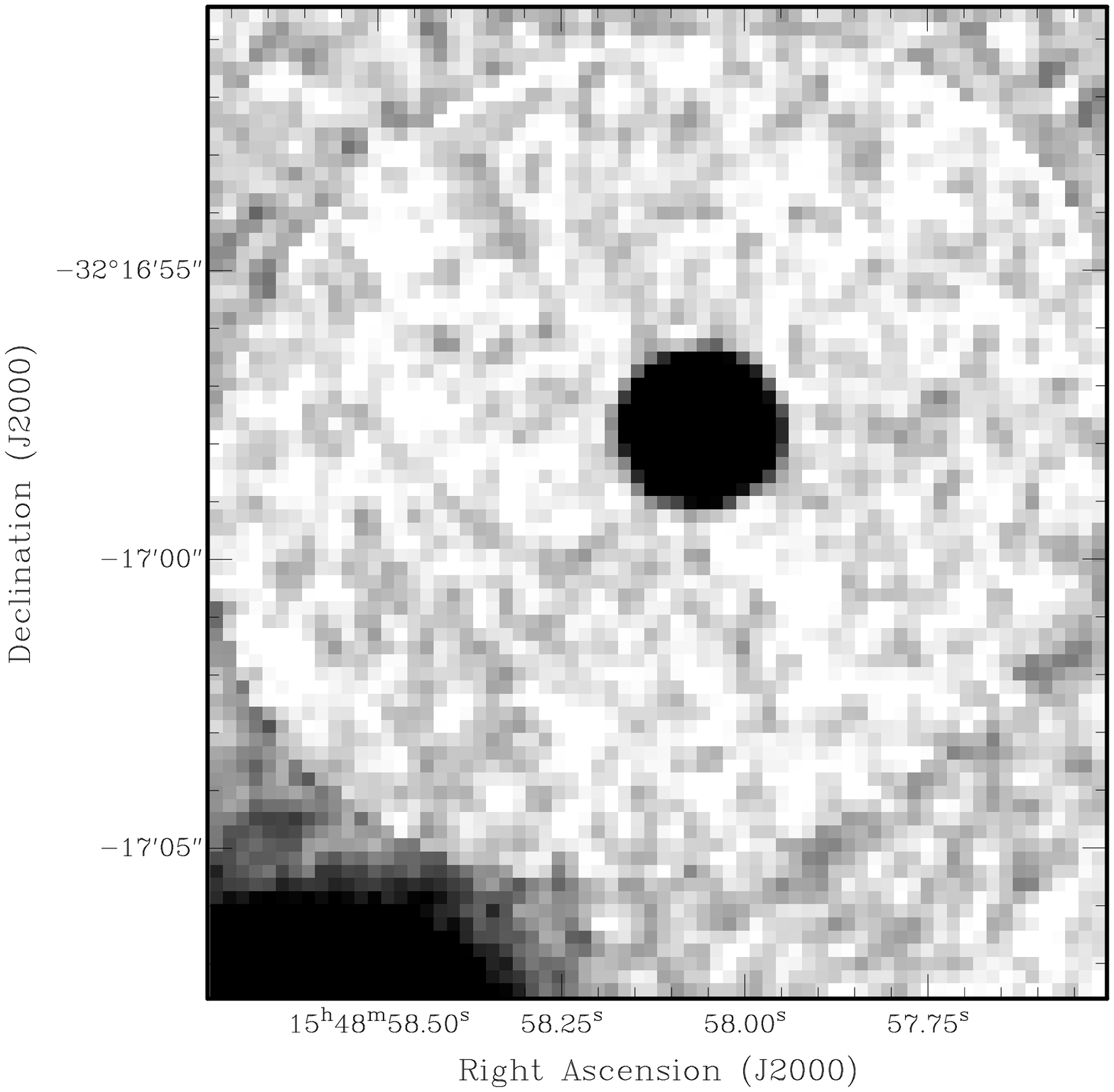}
\caption{\label{redwfires} Residual R-band AAT WFI image of the host galaxy 
after subtracting the model galaxy. The central 2 arcseconds are not included in 
the model }
\end{figure*}
\begin{figure*}
\includegraphics[angle=-90,width=10cm]{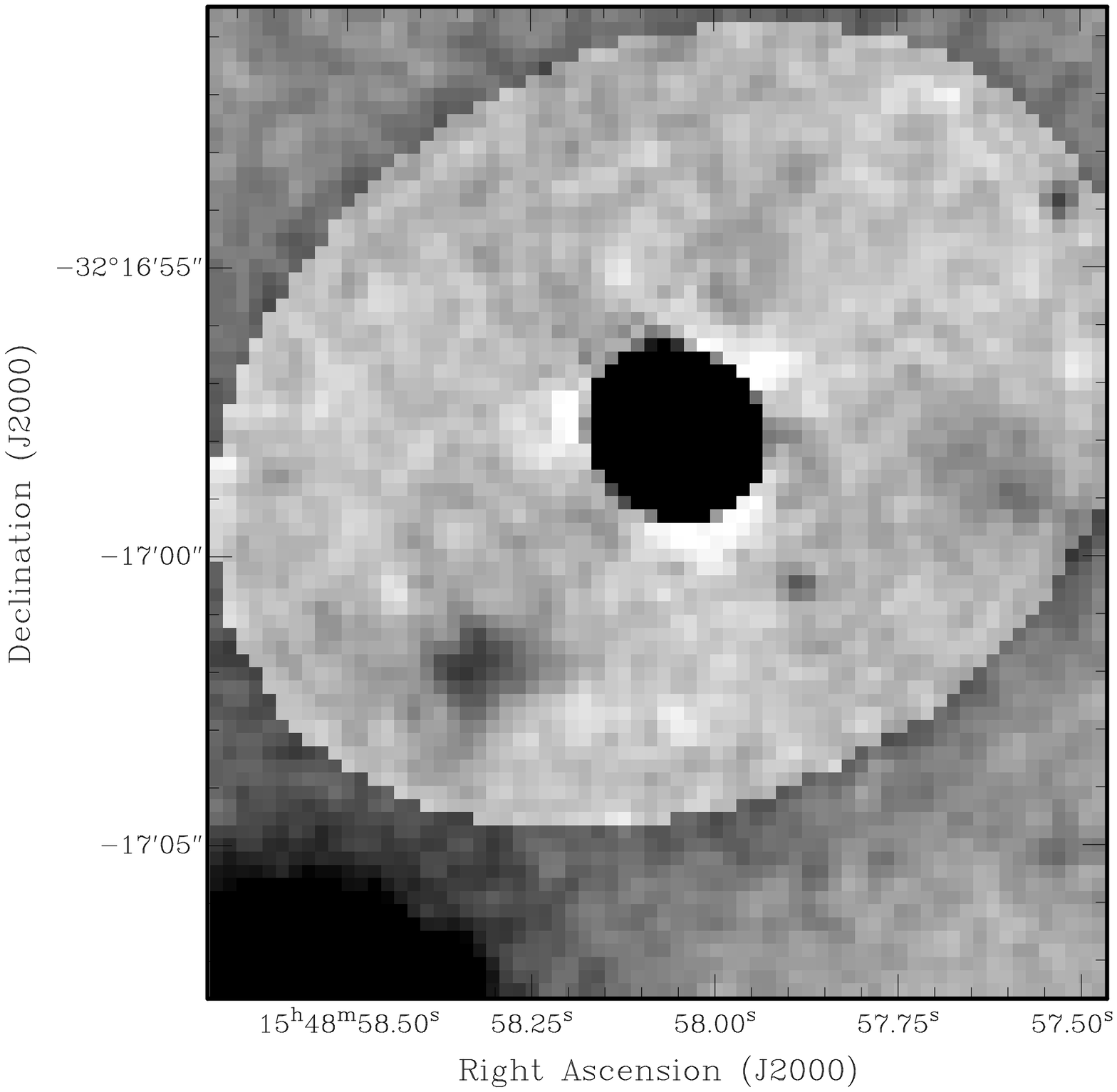}
\caption{\label{bluewfires} Residual B-band AAT WFI image of the host galaxy 
after subtracting the model galaxy. The central 2 arcseconds are not included in 
the model. }
\end{figure*}

\begin{figure*}
\includegraphics[height=\textwidth, angle=0]{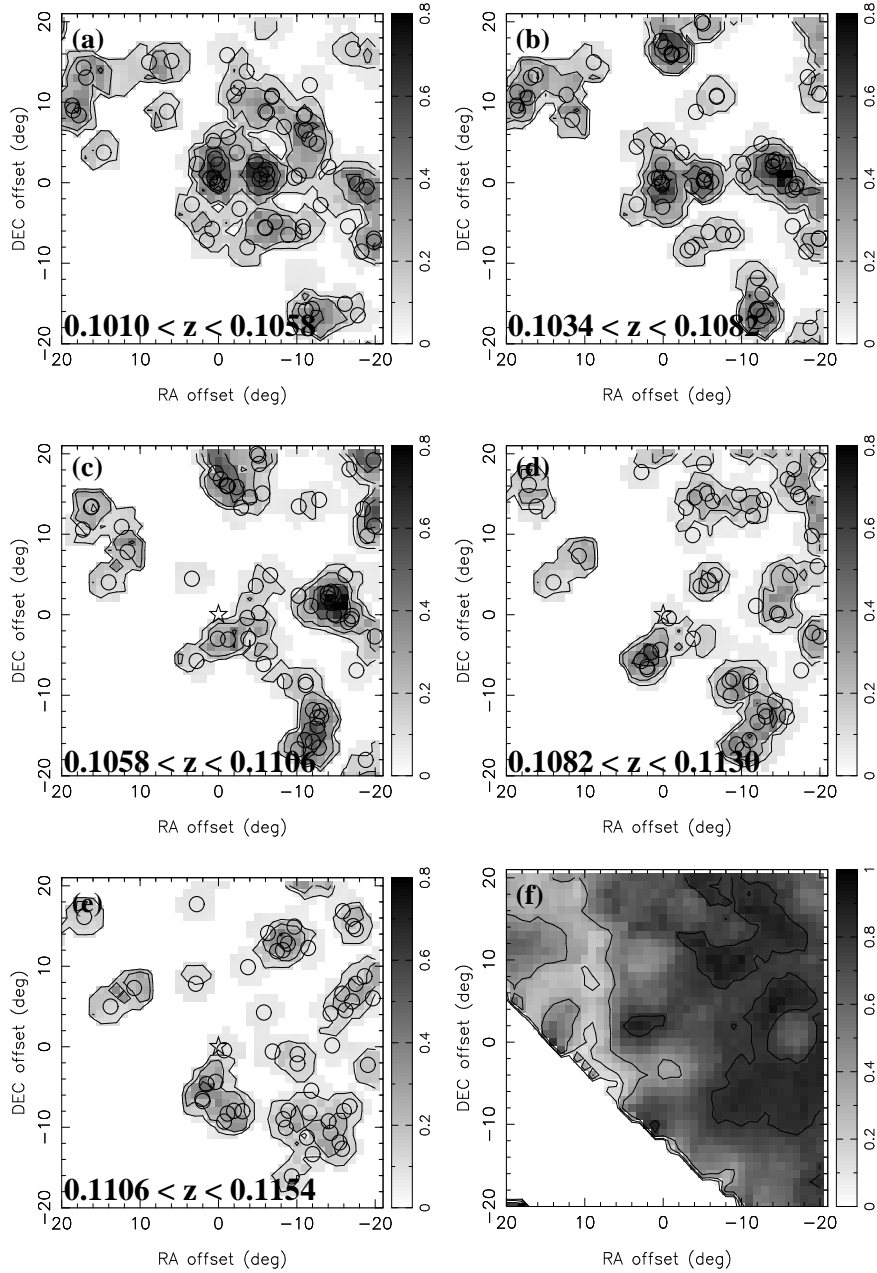}
\caption{\label{6df} 
Panels (a) through (e) show the smoothed 6dF galaxy density (per sq. degree) in 
the vicinity of the host galaxy of \pksns. Panels are 20 Mpc deep in redshift 
space and overlap by 10 Mpc. Contour levels are at: 0.1, 0.2, 0.4 and 0.8. The 
open circles mark the locations of 6dF galaxies in the corresponding redshift 
slices. The position of the host galaxy is indicated with a star symbol. Panel 
(f) shows the redshift completeness of the 6dFGS.
}
\end{figure*}

\begin{figure*}
\includegraphics[height=10cm]{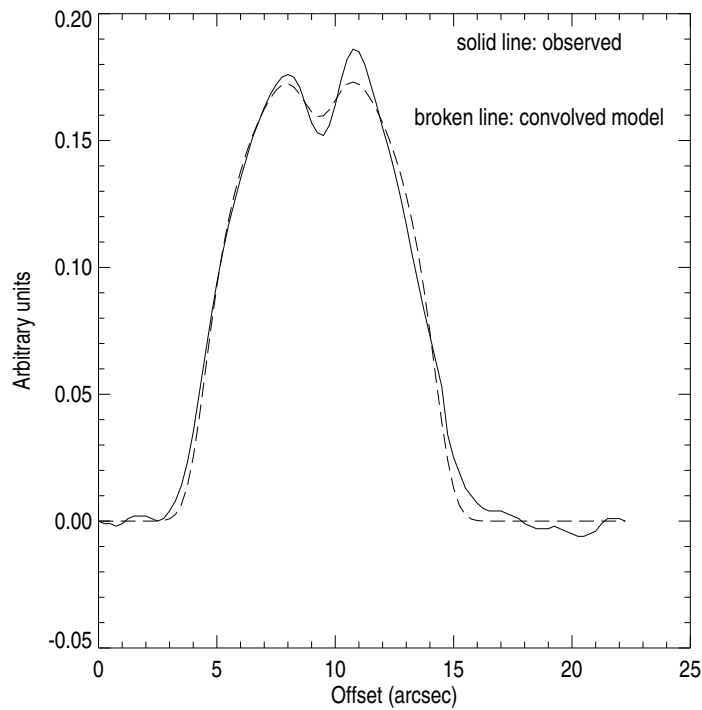}
\caption{\label{profile} Observed mean profile across the width of the southern 
inner source.}
\end{figure*}

\begin{figure*}
\includegraphics[height=10cm]{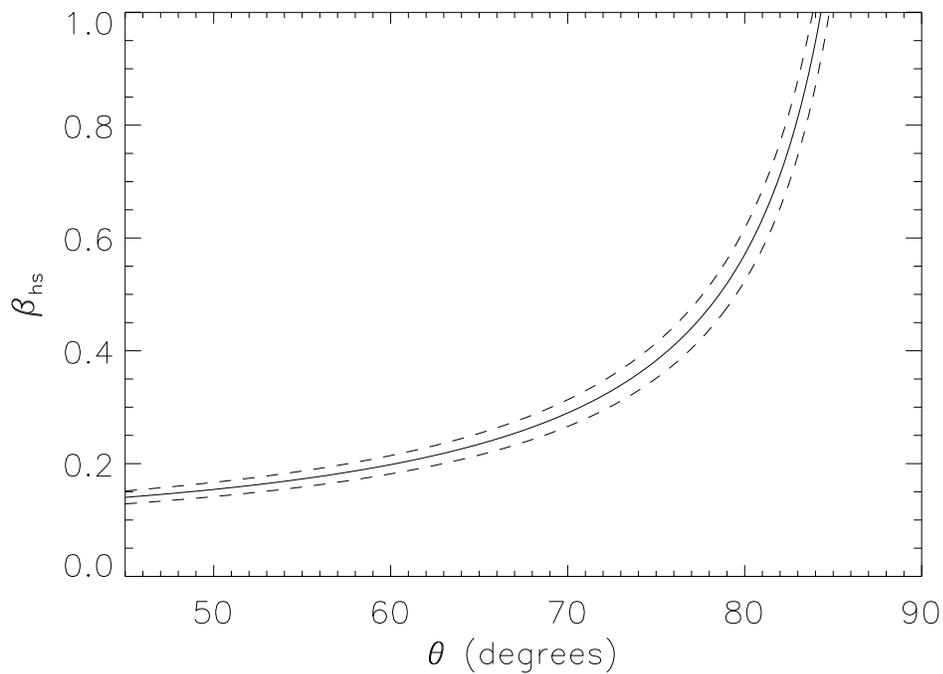}
\caption{\label{betahs} Hot spot advance velocity $\beta_{\rm hs}$ as a function 
of inclination angle
$\theta$, implied by the arm-length asymmetry ratio of $D=1.22\pm0.02$. The 
dashed curves show the lower and upper limits respectively implied by the 
uncertainty in the arm-length asymmetry measurement.}
\end{figure*}

\begin{figure*}
\includegraphics[height=10cm]{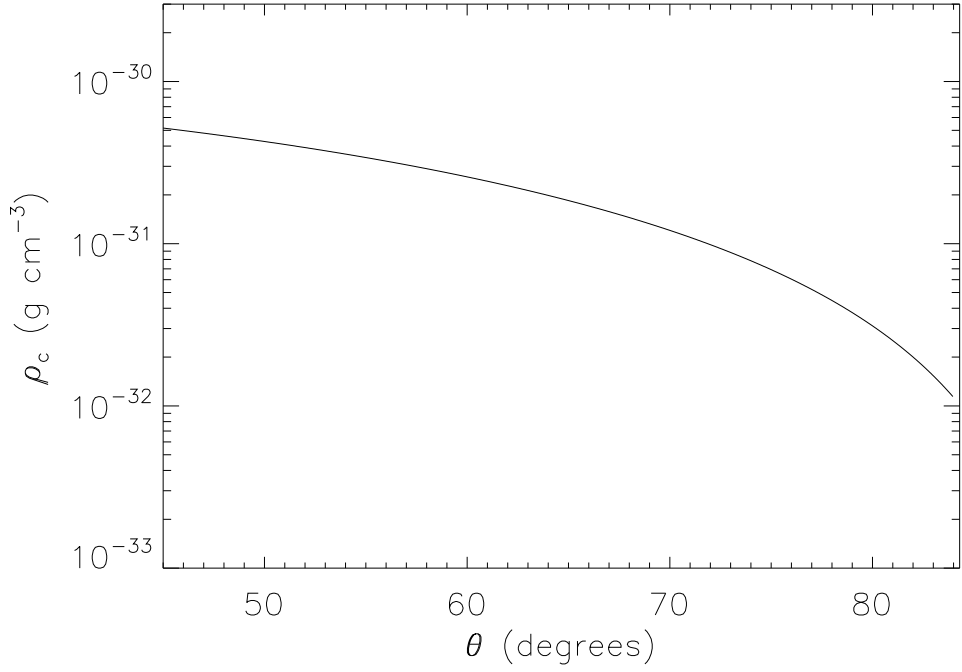}
\caption{\label{rho_theta}  Density in the outer lobes $\rho_{\rm c}$ as a 
function of  inclination angle $\theta$.}
\end{figure*}

\begin{figure*}
\includegraphics[height=10cm]{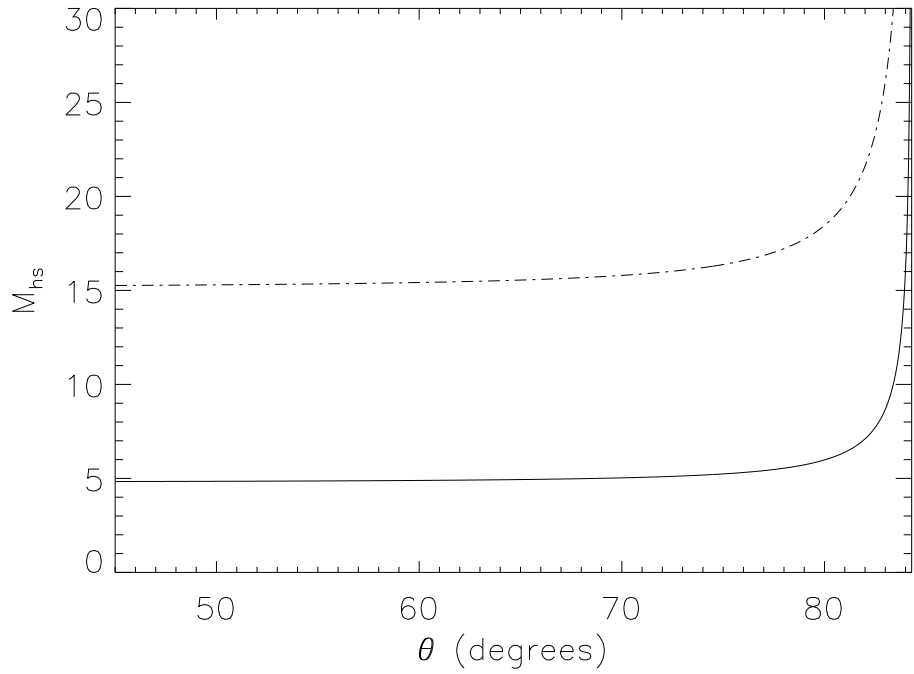}
\caption{\label{mach_theta} The external Mach number of the hot spot $M_
{\rm hs}$ 
as a function of inclination angle $\theta$ for $p_{\rm hs} = p_{\rm min}$ 
(solid line) and $10p_{\rm min}$ (dot-dashed line).}
\end{figure*}

\begin{figure*}
\includegraphics[height=10cm]{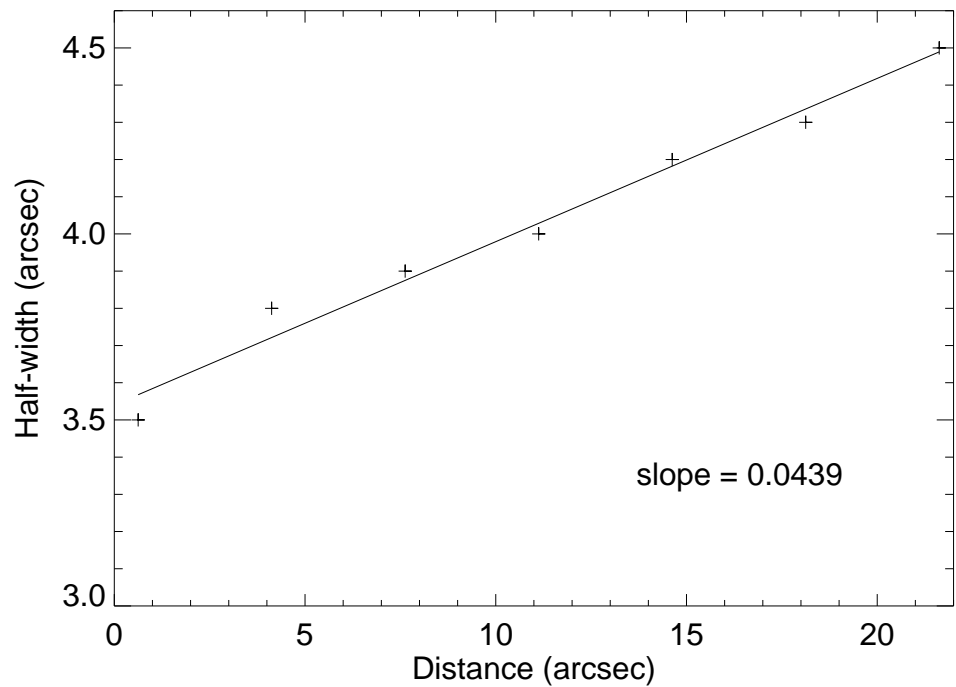}
\caption{\label{f:openang} Plotted widths of profiles of the SE trail.}
\end{figure*}

\begin{figure*}
\includegraphics[height=10cm]{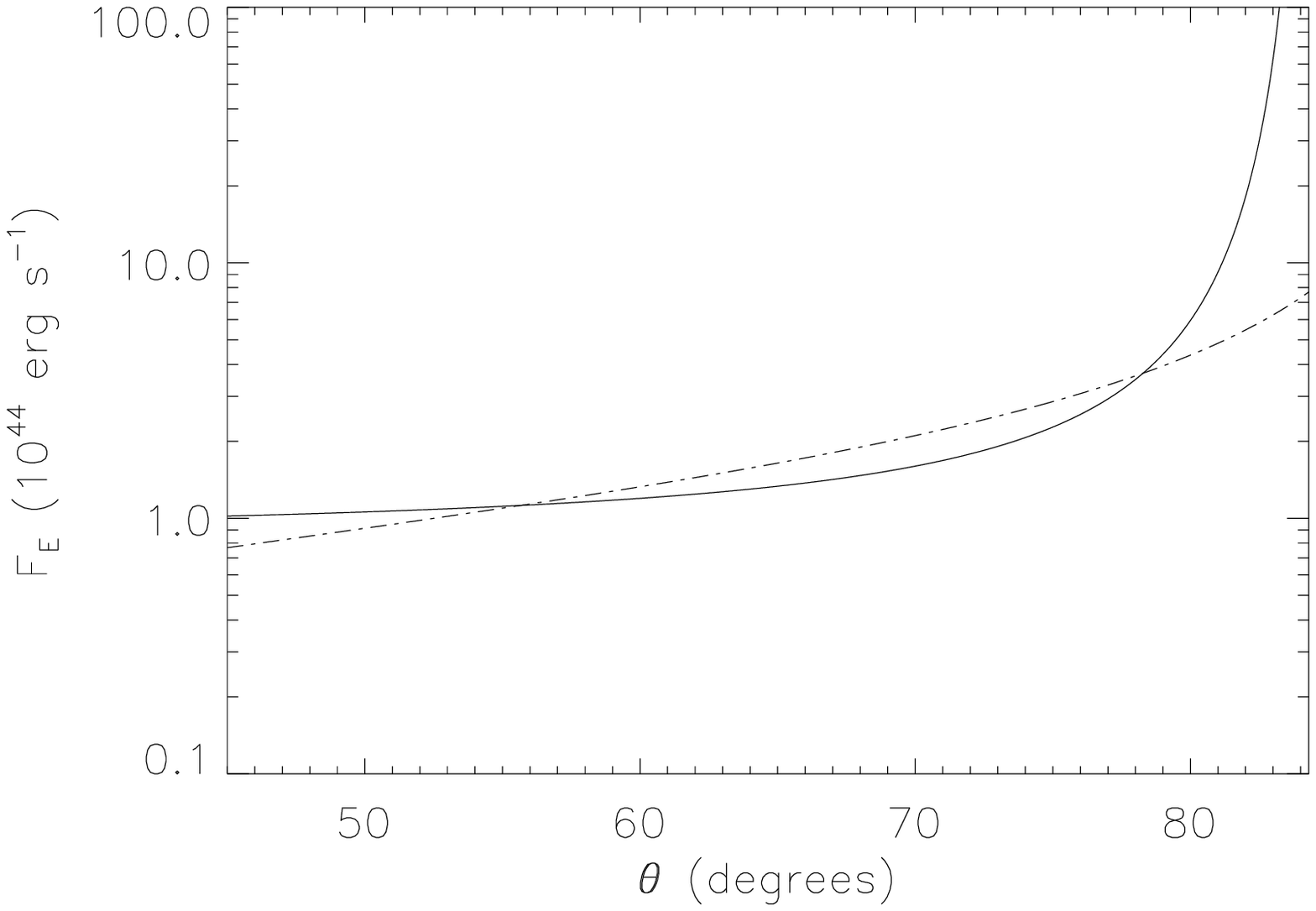}
\caption{\label{fe_theta} The energy flux of the restarting jets $F_{\rm E}$ as 
a function of inclination angle $\theta$, computed from equation~\ref{e:fes}  
(solid line) and equation~\ref{e:fes2} (dot-dashed line).}
\end{figure*}

\appendix

\section{Imaging combined VLA and ATCA datasets}
\label{s:appendix0}

Our VLA and ATCA 22 cm visibilities were combined and imaged using mosaic imaging techniques. The observations made by the individual telescopes were not mosaic observations, in which visibility data corresponding to multiple antenna pointings are acquired. Rather, the VLA and ATCA observations were both single pointing observations, and both observations used the same antenna pointing position. Nevertheless, it is necessary to treat the data specially while imaging because the antenna sizes differ in the two telescopes and, consequently the primary beams are different. Therefore, the sky source represented in the visibility data differ in the two telescopes, and sources offset from the pointing position have different primary beam attenuations. 
\par  
The combined VLA and ATCA 22 cm image of  \pks  shown in Figure~\ref{vlaatca}, was computed in three separate steps, using the mosaic imaging, deconvolution and restoration tasks available in the MIRIAD package. Firstly, a linear mosaic of the 22 cm VLA and ATCA visibilities was performed using task INVERT. After converting the ATCA linear and VLA circular polarization measurements into Stokes parameters, the routine images
the two pointings (visibility datasets) separately and then combines the resulting dirty images in a linear mosaic process. Pixels in the individual images are weighted to correct for the primary beam attenuation, as well as to minimize the noise in the combined image. Specifically, the resultant dirty image that is output by INVERT, $I_(\ell,m)$, is given by
\begin{equation}
I(\ell,m) =W(\ell,m) \frac{\sum_{i} P_{i}(\ell-\ell_{i},m-m_{i})I_{i}(\ell,m)/\sigma^{2}_{i}}
{\sum_{i}P_{i}^{2}(\ell-\ell_{i},m-m_{i})/\sigma^{2}_{i}}
\end{equation}
\citep{sault96} where the summation, $i$, is over the pointing centers
$(\ell_{i},m_{i})$. The parameter $I_{i}(\ell,m)$ refers to the image formed from the $i$'th pointing and $P_{i}(\ell,m)$ is the primary beam pattern for the corresponding telescope.  $W(\ell, m)$ is a weighting factor, which ensures that the noise at each mosaic pixel does not exceed a particular threshold value. 
In the case considered here, the pointing centers refer to the individual ATCA and VLA observations and the weighting function, $W(\ell,m) = 1$, within the region of the mosaic containing \pksns. Therefore, the attenuation due to the VLA and ATCA primary beams is totally corrected for over this area. 
\par
In addition to creating the combined dirty image, INVERT outputs a data cube containing the dirty sythesized beam patterns and primary beam models for each pointing. This enables the true point spread function, which is essentially a linear mosaic of the individual synthesized beam patterns, to be computed at any position in the image during the deconvolution stage. 
\par
In the second step, the MIRIAD routine MOSMEM \citep{sault96} was used to perform a joint maximum entropy deconvolution of the 22 cm mosaiced dirty image. For each pointing, the routine multiplies the relevant  primary beam with a prospective global model of the sky and then convolves the product with the dirty beam for that pointing. The dirty images are then linearly mosaiced together, and subtracted from the combined dirty image, to form the residuals.  Through this process, the position-variant point spread function of the linear mosaic is accounted for.
\par
In the final step, the global sky model and deconvolution residuals, as determined by MOSMEM, were input to the MIRIAD task RESTOR to compute a clean image. 
RESTOR convolves the global model with a constant Gaussian CLEAN beam (a beam that is not a function of position) and then folds the residuals into the result. The actual point spread function was calculated at various positions in the linear mosaic; these were found to be fairly constant with parameters similar to that of the synthesized VLA beam. Therefore, we let RESTOR follow the default approach, which is to convolve the model with a CLEAN beam whose parameters are determined by fitting to the synthesized beam of the first pointing; this was the VLA pointing in our case.

\section{Dissipation timescale for a hot spot}
\label{s:appendix}

If the beams from the central engine stop abruptly, and the last jet material 
passes through the hot spots at the ends of radio lobes, the hot spots will 
dissipate into the surrounding lobe. In this section we estimate the dissipation 
timescale assuming that the hot spots are spherically symmetric and the 
expansion is adiabatic.
\par
Let the initial and final radii, pressures and rest mass particle densities within 
the hot spots be $R_{0}$ and $R$, $p_{0}$, $p_{hs}$, $\rho_{0}$ and $\rho$,  
respectively. The thermal particle density in the external lobe is denoted by $
\rho_{\rm ext}$. Assuming that the expansion of the hot spot, with speed $v_{\rm 
exp}$, is ram pressure confined as a result of the entrained thermal gas in the 
external lobe,
\begin{equation}
v_{\rm exp} = \left(\frac{p_{hs}}{\rho_{\rm ext}}\right)^{1/2}.
\end{equation}
Since $\rho R^{3}$ is a constant in the expansion, which we assume is adiabatic,
\begin{equation}
\frac{p_{\rm hs}}{p_{0}} = \left( \frac{\rho}{\rho_{0}} \right)^{\gamma},
\end{equation}
where $\gamma = 4/3$ is the adiabatic index of the relativistic gas within the 
expanding hot spot. It follows that
\begin{equation}
\frac{p_{\rm hs}}{p_{0}} =  \left( \frac{R}{R_{0}} \right)^{-4}.
\label{e:app}
 \end{equation}
 The expansion velocity is then
 \begin{equation}
 \frac{dR}{dt} = \left(\frac{p_{hs}}{\rho_{\rm ext}}\right)^{1/2} = \left(\frac
{p_{0}}{\rho_{\rm ext}}\right)^{1/2}
 \left( \frac{R}{R_{0}}\right)^{-2}
 \end{equation}
 Denoting  $R/R_{0}$ by $y$,
 \begin{equation}
 \frac{dy}{dt} = \left(\frac{p_{0}}{\rho_{\rm ext}R_{0}^2}\right)^{1/2} y^{-2}.
 \end{equation}
This equation has the solution,
\begin{equation}
y = \left[ 1+3 \left( \frac{p_{0}}{\rho_{\rm ext}R_{0}^2}\right)^{1/2} t \right]
^{1/3},
\end{equation}
which may be rearranged to give an expression for the expansion time $t$:
\begin{equation}
t = \frac{1}{3}\left[ \left(\frac{p}{p_{0}}\right)^{-3/4}-1\right] \left( \frac
{\rho_{\rm ext} R_{0}^2}{p_{0}}.
\right)^{1/2}
\end{equation}

\section{Estimate of jet energy flux, for a restarted jet, from hot spot 
parameters}
\label{s:appendix2}

In this section we use the momentum balance at the end of the hot spot to 
express the jet energy flux in terms of the observable parameters of the hot 
spot and the density of the cocoon.

As in the body of this paper, we let $\beta_{\rm jet}$ be the ratio of jet speed 
to the speed of light as measured in the rest frame of the host galaxy, $\Gamma_
{\rm jet}$ is the corresponding Lorentz factor, $A_{\rm jet}$ is the cross-
sectional area of the jet. $\beta_{\rm hs}$, $\Gamma_{\rm hs}$ and $A_{\rm hs}$ 
are the corresponding parameters for the hot spot. $\rho_{\rm c}$ is the density 
of the cocoon surrounding the restarted jet.  We also allow for the fact that 
near the working surface the momentum flux  may be spread over a wider area than 
the instantaneous jet cross-section either as a result of the jet direction 
changing or the flow dynamics in the vicinity of the working surface.Hence the 
cross-sectional area of the hot spot $A_{\rm hs} \geq A_{\rm jet}$.

We denote the jet speed and corresponding Lorentz factor in the frame of the 
contact discontinuity, or hot spot, by additional subscripts: `hs'. In the frame 
of the hot spot, momentum balance gives:
\begin{equation}
\Gamma^2_{\rm jet,hs} \beta_{\rm jet,hs}^2 (\rho_{\rm jet} c^2 + 4 p_{\rm jet})
A_{\rm jet} = \rho_c c^2 \beta_{\rm hs}^2 A_{\rm hs}.
\end{equation} 
Lorentz transformation yields a relationship between parameters in the frame of 
the host galaxy to that in the frame of the hot spot:
\begin{equation}
\Gamma_{\rm jet,hs}^2 \beta_{\rm jet,hs}^2  
= \Gamma_{\rm hs}^2 \Gamma_{\rm jet}^2 (\beta_{\rm jet} - \beta_{\rm hs})^2.
\end{equation}
and the momentum balance equation reads
\begin{equation}
4 \Gamma_{\rm hs}^2 \Gamma_{\rm jet}^2 p_{\rm jet} (1+\chi_{\rm jet}) (\beta_
{\rm jet} - \beta_{\rm hs})^2
A_{\rm jet} = \rho_c c^2 \beta_{\rm hs}^2 A_{\rm hs}
\label{e:fb}
\end{equation}
where $\chi_{\rm jet} = \rho_{\rm jet} c^2/ 4 p_{\rm jet}$.

The energy flux in the relativistic beam is given by:
\begin{equation}
F_{\rm E} = 4 c \Gamma_{\rm jet}^2 \beta_{\rm jet} 
p_{\rm jet} \left(1 + \frac{\Gamma_{\rm jet}-1}{\Gamma_{\rm jet}} \chi_{\rm jet} 
\right)
A_{\rm jet}. 
\label{e:fe}
\end{equation}
Dividing this equation by equation~(\ref{e:fb}) gives the following expression 
for the energy flux:
\begin{equation}
F_{\rm E} = \left[ \frac {1 + 
\frac{\Gamma_{\rm jet}-1}{\Gamma_{\rm jet}} \chi_{\rm jet}}{1 + \chi_{\rm jet}}
\right]
\times \rho_{\rm c} c^3 \beta_{\rm jet} \Gamma_{\rm hs}^{-2}
\frac {(\beta_{\rm hs}/\beta_{\rm jet})^2}{[1 - (\beta_{\rm hs}/\beta_{\rm 
jet})]^2}
A_{\rm hs}.
\end{equation}
For even modest Lorentz factors for the jet, the leading factor in square 
brackets is close to unity, irrespective of the value of $\chi_{\rm jet}$. 
Additionally, $\beta_{\rm jet}$ is close to unity.  Hence this expression may be 
used to estimate the jet energy flux in terms of the hot spot speed, hot spot 
area and cocoon density.


\bsp

\label{lastpage}

\end{document}